%% file: hintofisocurvature-14.tex
\documentclass[12pt,letterpaper,english,aps,tightenlinesletterpaper,groupaddress,secnumarabic,nofootinbib]{revtex4}
\usepackage{mathptmx}

\usepackage[T1]{fontenc}
\usepackage[latin9]{inputenc}
\usepackage{babel}
\usepackage{amsmath}
\usepackage{amssymb}
\usepackage{graphicx}
\usepackage{esint}
\usepackage[unicode=true,pdfusetitle,
 bookmarks=true,bookmarksnumbered=false,bookmarksopen=false,
 breaklinks=false,pdfborder={0 0 1},backref=section,colorlinks=false]
 {hyperref}
\hypersetup{
 citecolor=blue}

\makeatletter

\pdfpageheight\paperheight
\pdfpagewidth\paperwidth

\providecommand{\tabularnewline}{\\}

\@ifundefined{textcolor}{}
{%
 \definecolor{BLACK}{gray}{0}
 \definecolor{WHITE}{gray}{1}
 \definecolor{RED}{rgb}{1,0,0}
 \definecolor{GREEN}{rgb}{0,1,0}
 \definecolor{BLUE}{rgb}{0,0,1}
 \definecolor{CYAN}{cmyk}{1,0,0,0}
 \definecolor{MAGENTA}{cmyk}{0,1,0,0}
 \definecolor{YELLOW}{cmyk}{0,0,1,0}
}

\usepackage{babel}

\usepackage{amsfonts}
\usepackage{bbm}
\usepackage{euscript}
\usepackage{dcolumn}
\usepackage{bm}
\usepackage{epsfig}\epsfclipon
\usepackage{slashed}

\newcommand{\Camb}{{\tt{CAMB}}}
\newcommand{\redtime}{{\tt{redTime}}}

\newcommand{\kappastar}{\kappa_\star}

\newcommand{\omegab}{\omega_\mathrm{b}}
\newcommand{\omegac}{\omega_\mathrm{c}}

\usepackage{color}

\makeatother

\begin{document}

\title{A Hint of a Blue Axion Isocurvature Spectrum?}

\author{Daniel J. H. Chung}

\email{danielchung@wisc.edu}

\affiliation{Department of Physics, University of Wisconsin-Madison, Madison,
WI 53706, USA}

\author{Amol Upadhye}

\email{aupadhye@wisc.edu}

\affiliation{Department of Physics, University of Wisconsin-Madison, Madison,
WI 53706, USA}
\begin{abstract}
It is known that if the Peccei-Quinn symmetry breaking field is displaced
from its minimum during inflation, the axion isocurvature spectrum
is generically strongly blue tilted with a break transition to a flat
spectrum. We fit this spectrum (incorporated into the ``vanilla''
$\Lambda$-CDM cosmological model) to the Planck and BOSS DR11 data
and find a mild hint for the presence of axionic blue-tilted isocurvature
perturbations. We find the best fit parameter region is consistent
with all of the dark matter being composed of QCD axions in the context
of inflationary cosmology with an expansion rate of order $10^{8}$
GeV, the axion decay constant of order $10^{13}$ GeV, and the initial
misalignment angle of order unity. Intriguingly, isocurvature with
a spectral break may at least partially explain the low-$\ell$ vs.~high-$\ell$
anomalies seen in the CMB data. 
\end{abstract}
\maketitle

\section{Introduction}

QCD axions \cite{Peccei:1977hh,Weinberg:1977ma,Wilczek:1977pj,Kim:1979if,Shifman:1979if,Zhitnitsky:1980tq,Dine:1981rt,Kim:2008hd}
are well motivated because they represent a simple elegant solution
to the strong CP problem and can be embedded in UV completions such
as string theory \cite{Svrcek:2006yi}. A huge literature exists regarding
the cosmological implications of the axions in which the field responsible
for Peccei-Quinn (PQ) symmetry breaking has not been displaced from
its minimum (see e.g.~\cite{Hertzberg:2008wr,Malik:2008im,Beltran:2006sq,Bucher:2004an,Fox:2004kb,Efstathiou:1986,Seckel:1985tj}).
In such cases, the isocurvature spectral index $n_{I}$ is very close
to $1$ which is often referred to as scale invariant. However, if
the PQ symmetry breaking field is displaced from its minimum during
inflation, blue spectral tilted isocurvature perturbations are naturally
generated \cite{Kasuya:2009up}. Indeed, the Goldstone theorem does
not apply in such cases because the axions do not represent perturbations
away from the vacuum \cite{Chung:2015pga}. Owing to the same physics
that governs the $\eta$-problem in inflation \cite{Dine:1983ys,Bertolami:1987xb,Copeland:1994vg,Dine:1995uk},
this class of models generically predicts an $n_{I}-1\sim O(1)$.
Furthermore, because the radial field eventually reaches its minimum,
this class of models generically predicts a break in the spectrum
where the spectral index transitions to that of scale invariance.
Indeed, since $n_{I}>$2.4 cannot be generated with a spectator scalar
field degree of freedom with a time independent mass \cite{Chung:2015tha},
a large spectral index in the context of inflationary cosmology predicts
a break in the spectrum for strongly blue-tilted isocurvature perturbations,
independently of the axion paradigm. This means interesting robust
information about the physics beyond the Standard Model of particle
physics (i.e.~the existence of a time dependent mass of a new particle)
can be gained from finding observational evidence for a strongly blue
tilted spectrum with a break. Because the axions are arguably the
best motivated underlying model for this class of scenarios producing
a break spectrum, we will call this strongly blue tilted isocurvature
spectrum with a break an axionic blue isocurvature (ABI) spectrum

The break region in the ABI spectrum cannot be computed analytically
using the standard techniques \cite{Chung:2015pga}. Recently, an
efficient 3-parameter fit function $\Delta_{S}^{2}(k,k_{\star},n_{I},\mathcal{Q}_{1})$
was constructed from generalizing a numerical investigation \cite{Chung:2016wvv}
of the model of \cite{Kasuya:2009up}, and this fit function has a
bump feature that can be numerically significant at an $O(1)$ level.
In this paper, we use this fit function in the context of $\Lambda$-CDM
cosmological model to fit 9 parameters to the PLANCK \cite{Adam_2016j,Aghanim_2016}
and Baryon Oscillation Spectroscopic Survey Data Release 11 (BOSS
DR11) \cite{Beutler_2014,Beutler_2014b}. We find that the data prefer
a non-zero ABI spectrum at the 1-sigma level with the best fit parameters
of about $\{k_{\star}/a_{0}=4.1_{-2.7}^{+14}\times10^{-2}{\rm Mpc}^{-1},\,\, n_{I}=2.76{}_{-0.59}^{+1.1},\,\, Q_{n}=0.96_{-0.93}^{+0.32}\}$
where $k_{\star}/a_{0}$ is the spectral location of the break, $n_{I}$
is the isocurvature spectral index, and $Q_{n}\times10^{-10}$ is
approximately the isocurvature power on BAO scales that can be compared
to $\Delta_{\zeta}^{2}\sim O(10^{-9})$ of the usual adiabatic perturbations.
This best fit region can be consistent with an initial axion angle
of $\theta_{+}(t_{i})=0.1$ and all of the dark matter being made
up of axions. For example, with this fiducial parameter choice, the
scale of inflation is given by the expansion rate during inflation
of $H\approx2\times10^{8}$ GeV and the axion decay constant of $F_{a}\sim10^{13}$
GeV. In this parameter range, the bump that was computed numerically
in \cite{Chung:2016wvv} contributes at the level of about 10\% for
the values of the fit parameters and changes the shape of the fit
contours only slightly. We also carried out a fit with $n_{I}=3.9$
and $k_{\star}/a_{0}=0.5/$~Mpc~ and find a $2\sigma$ preference
for a highly blue-tilted isocurvature with a power-law spectrum on
observable scales.

Since the smallest length scales probed by current CMB and galaxy
surveys are similar, we find the CMB data to be more constraining
due to their higher precision, though of course the two sets of observables
have different parametric degeneracies. There are no substantial tensions
between the two data sets; the most significant changes in the vanilla
parameters due to the BOSS data are the decreases in $\sigma_{8}$
and $\tau$ along their mutual degeneracy direction preserving $\sigma_{8}e^{-\tau}$
. In the isocurvature sector, we find that BOSS data increase the
preference for blue-tilted models with spectral breaks below observable
length scales.

The order of presentation will be as follows. In the next section,
we review the fitting function $\Delta_{S}^{2}(k,k_{\star},n_{I},\mathcal{Q}_{1})$
and a lamp-post model that inspired this. In Sec.~\ref{sec:Data-fit},
we present the ABI + $\Lambda$-CDM fit. In Sec.~\ref{sec:Lamp-post-model-interpretation},
we interpret the fit results in terms of the lamp-post model of \cite{Kasuya:2009up}.
We conclude with a summary of the work and speculations about this
work's connection to the low-$\ell$ and high-$\ell$ CMB data mismatch
noted in \cite{Addison:2015wyg}.

\section{\label{sec:A-brief-review}A brief review of the ABI spectrum parametrization}

Most of the axionic isocurvature literature focuses on the scenario
in which the Peccei-Quinn symmetry breaking field $f_{PQ}$ has already
relaxed to the minimum of the effective potential \cite{Hertzberg:2008wr,Malik:2008im,Beltran:2006sq,Bucher:2004an,Efstathiou:1986,Seckel:1985tj}.
However, in situations in which the radial direction has a mass of
order $H$, such an assumption is not well justified since the inflaton
itself is out of equilibrium during that time, and it may take many
efolds for $f_{PQ}$ to relax to the minimum of the effective potential
\cite{Kasuya:2009up}. In such cases, a strongly blue-tilted isocurvature
spectrum is generically generated. Particularly in supersymmetric
extensions of the Standard Model, flat directions abound, and $f_{PQ}$
realized as a flat direction field will generically have masses of
$O(H)$ \cite{Kasuya:2009up,Chung:2015pga} generating dynamics suitable
for the creation of ABI perturbations.%
\footnote{Indeed, this is a situation in which the $\eta$-problem of inflation
turns into an advantage.%
}

Although the ABI spectrum computed in \cite{Kasuya:2009up} is qualitatively
valid, it was noted in \cite{Chung:2015pga} that there is generically
a spectral gap in analytic computability (with the standard techniques)
surrounding the break region. In \cite{Chung:2016wvv}, we computed
numerically the ABI spectrum of the model analytically analyzed in
\cite{Kasuya:2009up,Chung:2015pga} and found that the spectrum indeed
has a nontrivial bump in the break region between the constant blue
tilt region and the scale invariant region with the transition spectral
width consistent with the predictions of \cite{Chung:2015pga}. The
ABI spectrum including the bump is fit well with the following function
defined by 3 parameters $k_{\star}$, $n_{I}$, and $\mathcal{Q}_{1}$:
\begin{equation}
\Delta_{S}^{2}(k,k_{\star},n_{I},\mathcal{Q}_{1})\approx\mathcal{Q}_{1}\frac{1+\alpha(n_{I})L\left[\frac{1}{\sigma(n_{I})}\ln\left(e^{-\mu(n_{I})}\frac{k}{k_{\star}}\right)\right]S\left[\frac{\lambda(n_{I})}{\sigma(n_{I})}\ln\left(e^{-\mu(n_{I})}\frac{k}{k_{\star}}\right)\right]}{\left[1+\left(\tilde{\rho}(n_{I})2^{2\sqrt{\frac{9}{4}-c_{+}(n_{I})}}\frac{\Gamma^{2}\left(\sqrt{\frac{9}{4}-c_{+}(n_{I})}\right)}{2\pi}\left(1+\frac{c_{+}(n_{I})}{0.9}\right)\left(\frac{k}{k_{\star}}\right)^{n_{I}-1}\right)^{-1/w}\right]^{w}}\label{eq:fitfunc}
\end{equation}
\begin{eqnarray}
L(x) & = & 1/(1+x^{2})\\
S(x) & = & 1+\tanh(x)
\end{eqnarray}
\begin{equation}
c_{+}(n_{I})=\frac{1}{4}(n_{I}-1)(7-n_{I}).\label{eq:cplusfunc}
\end{equation}
The parameters $\{\alpha,\sigma,\mu,\lambda,w,\tilde{\rho}\}$ are
numerical factors that can be deduced from an interpolation of a table
of numbers given in Table 1 of \cite{Chung:2016wvv}.

The broad features of the isocurvature power spectrum are described
by the large-scale spectral index $n_{I}$, the break position $k_{\star}$,
and the break width $w$. On top of this monotonic power spectrum
is a peak of height $\alpha$, width $\sigma$, position $\mu$, and
skewness $\lambda$, resulting from the axionic field sloshing around
the minimum of its potential during the spectral transition. For example,
for $n_{I}=3$, the parameter set is $\{\alpha=0.56,\,\,\sigma=0.46,\,\,\mu=0.126,\,\,\lambda=-0.035,\,\, w=0.84,\,\,\tilde{\rho}=1.2\}$.
The number $0.9$ in Eq.~(\ref{eq:fitfunc}) corresponds to making
a choice for a model dependent parameter that gives an approximate
fit to model-dependent numerically computed results when this number
is in the range $[0.5,1]$.

To test if the ABI spectrum shows up in the current data and to see
how it is constrained, we fit in Sec.~\ref{sec:Data-fit} the standard
six ``vanilla'' cosmological parameters ($\Lambda$-CDM) plus up
to three more parameters describing the ABI power spectrum. The standard
vanilla $\Lambda$-CDM parameters can be given as follows: 1) $n_{s}$,
the spectral index of adiabatic scalar perturbations; 2) $\sigma_{8}$,
the root-mean-squared power in $8$~Mpc$/h$ spheres where 3) $h=H_{0}/(100\mathrm{km/sec/Mpc})$,
the Hubble parameter; 4) $\omega_{{\rm c}}=\Omega_{\mathrm{c},0}h^{2}$,
where $\Omega_{\mathrm{c},0}$ is the density fraction of cold dark
matter (CDM) at the present time; 5) $\omega_{{\rm b}}=\Omega_{\mathrm{b},0}h^{2}$,
where $\Omega_{\mathrm{b},0}$ is the baryon density fraction; and
6) $\tau$, the optical depth to the cosmic microwave background.
Since neglecting the neutrino mass can lead to parameter biases, we
fix $\omega_{\nu}=\Omega_{\nu}h^{2}=0.0006$ for the fits unless specified
otherwise. For efficient Markov Chain Monte Carlo (MCMC) sampling
with a flat prior (i.e.~to minimize degeneracies), it is useful to
sample 
\begin{equation}
Q_{n}\equiv10^{10}\frac{\mathcal{Q}_{1}}{1+\left(\frac{k_{\star}}{k_{0}}\right)^{n_{I}}}\label{eq:Q1QNrelation}
\end{equation}
and 
\begin{equation}
\kappa_{\star}=\ln\frac{k_{\star}}{k_{0}}
\end{equation}
(where $k_{0}$ is a fiducial wave vector which we will set as $k_{0}/a_{0}=0.05\,\,{\rm Mpc}^{-1}$)
instead of the parameters $\mathcal{Q}_{1}$ and $k_{\star}$.

\section{\label{sec:Data-fit}Data fit}







In this section, we fit the mixed adiabatic-isocurvature cosmological
model presented in Sec.~\ref{sec:A-brief-review} to Planck and BOSS
data. We broadly classify three different ABI parameter regions as
follows: 
\begin{itemize}
\item KK: the model of Kasuya and Kawasaki, Ref.~\cite{Kasuya:2009up},
with the bump fit by Ref.~\cite{Chung:2016wvv}; 
\item NB: a no-bump version of KK, with the bump height $\alpha$ set to
zero and the break width parameter fixed to $w=1/3$; 
\item PWR: a simple power-law spectrum, which we implement by fixing $\kappa_{\star}=\ln(200)$
in the NB model. 
\end{itemize}
We are especially interested in models with the bluest tilts $n_{I}\lesssim4$
over much of the observable parameter space $k\sim k_{0}$. Hence
we also consider the following lamp-post models partially restricting
the allowed values of $n_{I}$ and $\kappa_{\star}$: 
\begin{itemize}
\item BLUE: the KK model with $n_{I}=3.9$; 
\item HI-BLUE: the BLUE model with the further restriction $\kappa_{\star}=\ln(10)$. 
\item LAMP-$N$: the KK model with $\kappa_{\star}=\ln(N)$; 
\end{itemize}
At small $N$, LAMP-$N$ approaches an ordinary flat isocurvature
model, while at large $N$ it approaches a power law. We will constrain
LAMP-$1$, LAMP-$2$, and LAMP-$10$.

\subsection{Analysis procedure}

\label{subsec:analysis_procedure}

The data analysis used here is that of Ref.~\cite{Upadhye_2017},
modified to include isocurvature perturbations. Briefly, we combine
the publicly-available Planck likelihood code of Refs.~\cite{Adam_2016j,Aghanim_2016}
with the BOSS DR11 redshift-space galaxy power spectrum of Refs.~\cite{Beutler_2014,Beutler_2014b},
and explore the likelihood using a Metropolis-Hastings Markov Chain
Monte Carlo algorithm with a broad set of priors given in Table~\ref{t:priors}.
We summarize this procedure here.

\begin{table}[tb]
\tabcolsep=0.095cm \begin{footnotesize} %
\begin{tabular}{c|c|c|c|c|c|c|c|c}
$n_{s}$  & $n_{I}$  & $\sigma_{8}$  & $Q_{n}$  & $h$  & $\omega_{{\rm c}}$  & $\omega_{{\rm b}}$  & $\tau$  & $\kappa_{\star}$ \tabularnewline
\hline 
$>0$  & $[1,3.94]$  & $>0$  & $>0$  & $[0.2,1]$  & $>0$  & $>0.001$  & $>0.01$  & $[-3.9,2.3]$\tabularnewline
\end{tabular}\end{footnotesize} \protect\protect\caption{ The prior probability distribution is uniform in the parameters $n_{s}$,
$n_{I}$, $\ln(\sigma_{8})$, $Q_{n}$, $\theta_{100}$, $\omegac$,
$\omegab$, $\tau$, and $\kappa_{\star}$, with the above bounds.
$\theta_{100}$, an approximation to the angular scale of acoustic
oscillations which is related to the parameter $h$, is described
in Sec.~III of Ref.~\cite{Upadhye_2017}. \label{t:priors} }
\end{table}

The Planck likelihood computation is divided into low-$\ell$ and
high-$\ell$ components. Since the low-$\ell$ polarization likelihood
is computationally expensive, we restrict our $\ell<30$ analysis
to the temperature power spectrum. For $\ell\geq30$ we employ the
simplified \texttt{plik-lite} function of Ref.~\cite{Aghanim_2016},
which we marginalize over the absolute calibration parameter $A_{\mathrm{Planck}}$
as recommended. CMB power spectra are computed using the \Camb~cosmology
code of Ref.~\cite{Lewis_2000} modified to include isocurvature
power spectra described by the fitting function of Ref.~\cite{Chung:2016wvv}
appropriate to models with blue-to-flat spectral breaks. For mixed
models combining adiabatic and isocurvature perturbations, we ran
\Camb~separately for adiabatic and isocurvature initial conditions,
then added to find the combined linear power spectra. Since CMB lensing
is a non-linear process, we also compiled a stand-alone version of
the \Camb~CMB lensing function, which we used to lens the combined
linear power spectra.

The BOSS DR11 analysis of Ref.~\cite{Beutler_2014} measures the
monopole and quadrupole of the redshift-space galaxy power spectrum
at an effective redshift of $z=0.57$. That reference provides the
window functions and covariance matrices necessary for comparing power
spectra to the BOSS data. Beginning with \Camb~inputs, we compute
the power spectra for mixed adiabatic and isocurvature models using
a modified version of the \redtime~non-linear redshift-space perturbation
code of Refs.~\cite{Upadhye_2016,Upadhye_2017}, based upon the Time-Renormalization
Group method of Ref.~\cite{Pietroni_2008}. Since the growth of large-scale
structure after decoupling is well-described by a single CDM+baryon
fluid, mixed initial conditions can easily be accommodated by adding
the linear adiabatic and isocurvature power spectra computed by \Camb~at
the redshift $z_{\mathrm{in}}$ at which the non-linear perturbative
computation is initialized. We choose $z_{\mathrm{in}}=200$, as tested
against N-body dark matter simulations in \cite{Upadhye_2014,Upadhye_2016}.

Galaxies are biased tracers of the underlying density field. Since
blue-tilted isocurvature changes the shape of the matter power spectrum,
we must accurately model the scale-dependence of galaxy bias. Reference~\cite{McDonald_Roy_2009}
describes a five-parameter model of galaxy bias, which is simplified
to a three-parameter model in Ref.~\cite{Saito_etal_2014}. We use
this three-parameter model unless otherwise noted. At each chain point,
we marginalize numerically over these bias parameters as in Ref.~\cite{Upadhye_2017}
in order to compute the likelihood.

MCMC convergence is tested using the potential scale reduction factor
$\sqrt{R}$, which approaches unity from above as the variance of
the means of several chains becomes much smaller than the mean of
the individual chain variances \cite{Gelman_Rubin_1992,Brooks_Gelman_1997}.
For each model and data combination, we run 5 chains, which we judge
to have converged when $\sqrt{R}<1.05$ for fixed $\omega_{\nu}$
and $\sqrt{R}<1.1$ for variable $\omega_{\nu}$; these are more stringent
than the convergence requirement $\sqrt{R}<1.2$ recommended in Ref.~\cite{Brooks_Gelman_1997}.




\subsection{CMB constraints}

\label{subsec:cmb_constraints}

{\renewcommand{\baselinestretch}{1.0}   \input{table2_cmb.tex}}

Marginalized constraints on the vanilla and isocurvature parameters
from Planck data alone are shown in Table~\ref{t:constraints1d_cmb}.
Since Refs.~\cite{Ade_2016m,Ade_2016u} caution that low-level $T\rightarrow E$
leakage may contaminate the polarization data in a way which mimics
isocurvature, we begin by evaluating the effects of such leakage on
our parameter constraints. The first two columns of Table~\ref{t:constraints1d_cmb}
compare constraints using $C_{\ell}^{TT}$ only to those using $C_{\ell}^{TT}$,
$C_{\ell}^{TE}$, and $C_{\ell}^{EE}$. Since all parameter shifts
are substantially less than $1\sigma$, we conclude that the isocurvature
model considered here is insensitive to any residual $T\rightarrow E$
leakage. Henceforth we use Planck temperature and polarization data.

Comparing the vanilla parameters in Table~\ref{t:constraints1d_cmb}
to those in Table~3 of Ref.~\cite{Ade_2016m}, we see that parameter
shifts are less than $0.6\sigma$ except for $\sigma_{8}$ and $\tau$,
which both increase by $\approx1\sigma$ when isocurvature is included.
However, these increase together along their mutual degeneracy direction.
Since the CMB constrains the combination $\sigma_{8}\exp(-\tau)$
more tightly than either of these parameters individually, we expect
$\sigma_{8}$ and $\tau$ to change in such a way that $\Delta\sigma_{8}/\sigma_{8}\approx\Delta\tau$,
which is consistent with the shifts seen in Table~\ref{t:constraints1d_cmb}.

\begin{figure}[tb]
\begin{centering}
\includegraphics[width=2.18in]{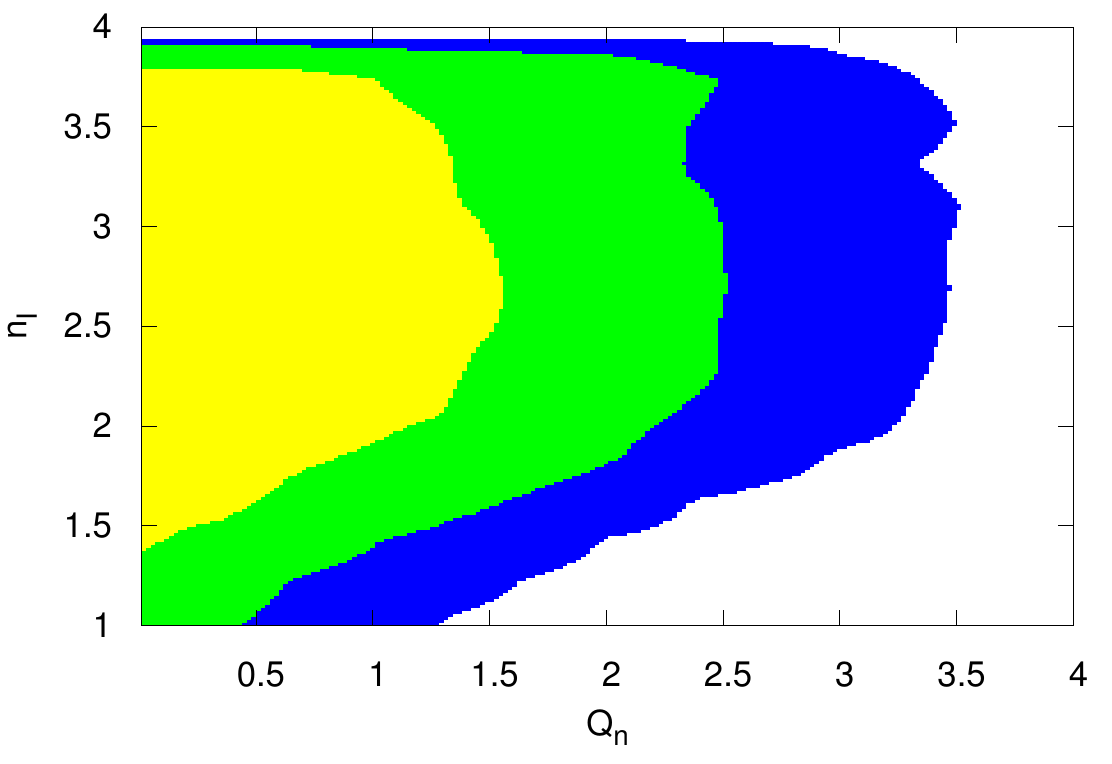} \includegraphics[width=2.1in]{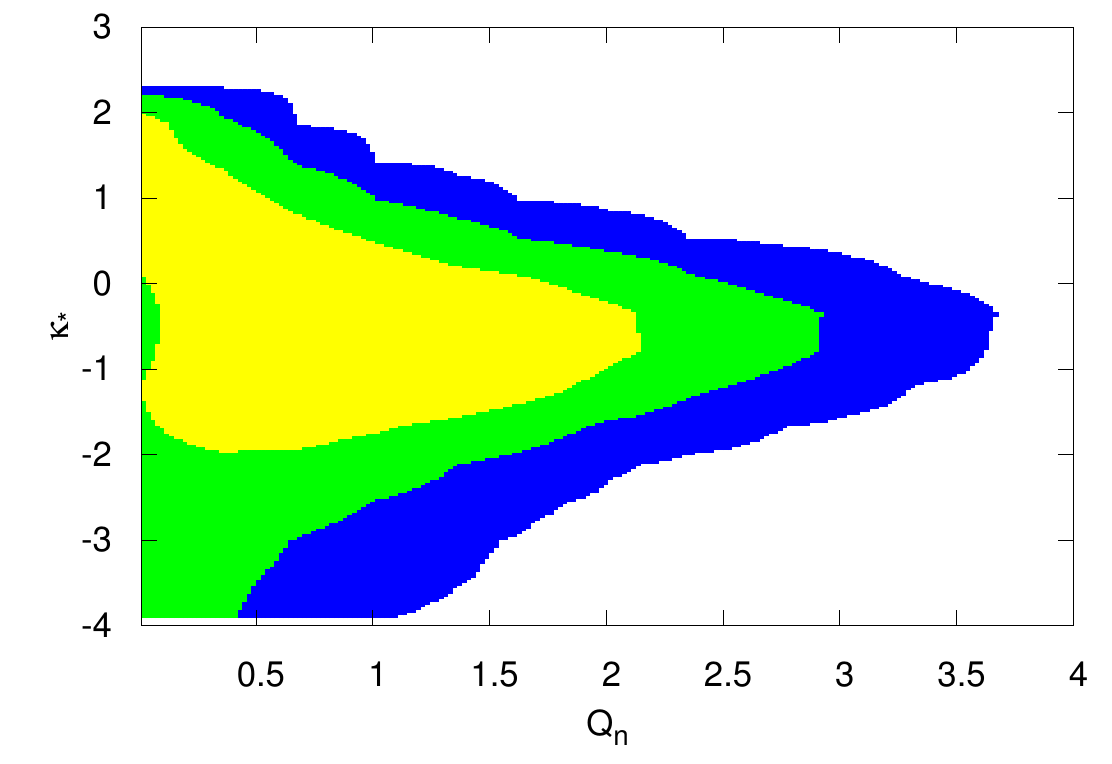}
\includegraphics[width=2.1in]{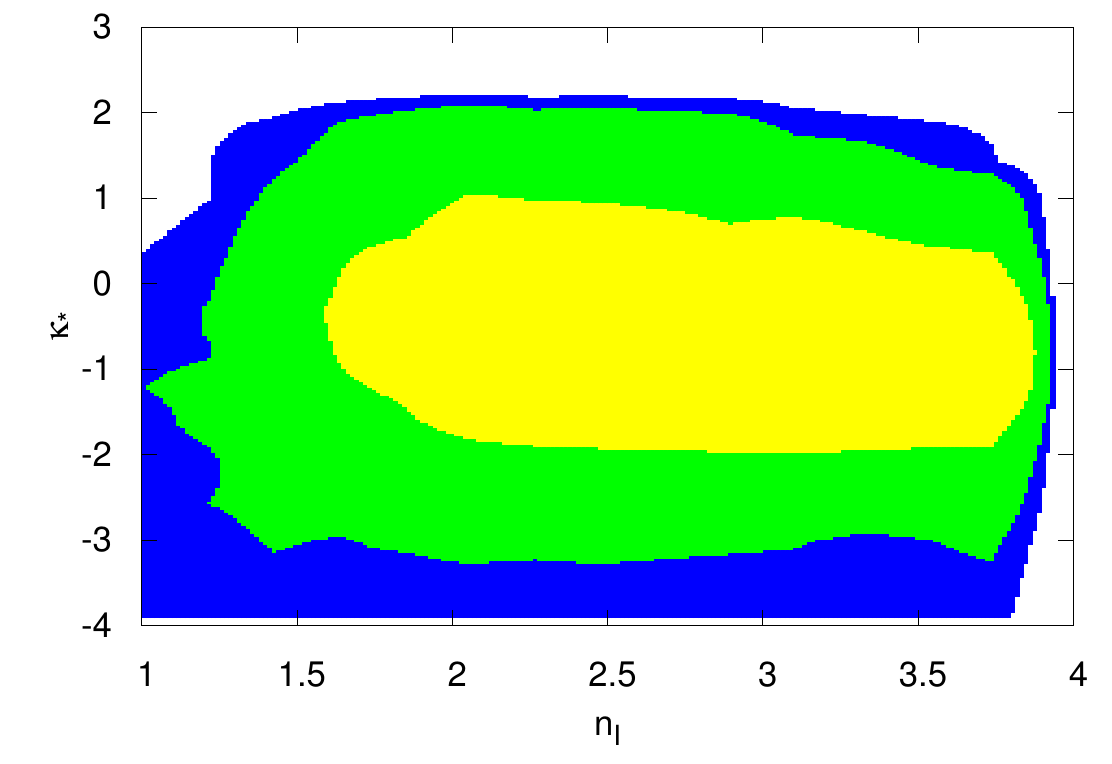}
\par\end{centering}

\centering{}\protect\caption{ Constraints on the KK model using Planck data. Light (yellow), medium
(green), and dark (blue) shaded regions identify $68\%$, $95\%$,
and $99.7\%$ confidence contours, respectively. \label{f:KK_P_iso_constraints} }
\end{figure}

Since $Q_{n}=0$ is allowed at the $1\sigma$ level for the KK and
NB models, and slightly more than $1\sigma$ for the PWR model, we
conclude that Planck data alone do not significantly prefer any of
the isocurvature models in the table. While there is a slight preference
for $n_{I}\approx2.7$ and $\kappastar\approx-0.5$, the $95\%$ allowed
regions for both of these parameters include nearly the full ranges
$1\leq n_{I}\leq3.94$ and $\ln(1/50)\leq\kappastar\leq\ln(10)$.
Figure~\ref{f:KK_P_iso_constraints} shows marginalized two-dimensional
constraints on the isocurvature parameters in the KK model. Note that
for the smallest $\kappastar$ values, the isocurvature spectrum is
flat over most of the observable range, meaning that $n_{I}$ is poorly
constrained.


\subsection{Galaxy survey constraints}

\label{subsec:galaxy_survey_constraints}

{\renewcommand{\baselinestretch}{1.0}   \input{table3_cmb+gal.tex}}

Next we combine the BOSS DR11 galaxy survey data with the Planck data.
We begin by testing the robustness of our constraints with respect
to the inclusion of additional parameters describing the scale-dependent
bias. The first three columns of Table~\ref{t:constraints1d_cmb+gal}
constrain the KK model using Planck and BOSS data, marginalizing over
the $5$-parameter bias model of Ref.~\cite{McDonald_Roy_2009} for
the first two columns and $3$-parameter bias model for the other
columns. Comparing the first and the third column, the constraints
on $h$ and $\omega_{\mathrm{c}}$ shift by $\approx0.3\sigma$, while
all remaining parameters shift by less than $0.15\sigma$, and the
isocurvature parameters by $\leq0.03\sigma$, suggesting that the
$3$-parameter bias model used henceforth (unless specified otherwise)
provides robust constraints. Note that although allowing variations
in the sum of the neutrino masses leads to an increase in the best
fit value of $\kappa_{*}$ as can be seen in the second column, the
shift is statistically insignificant since it is much smaller than
a $1\sigma$ variation. 

\begin{figure}[tb]
\begin{centering}
\includegraphics[width=2.18in]{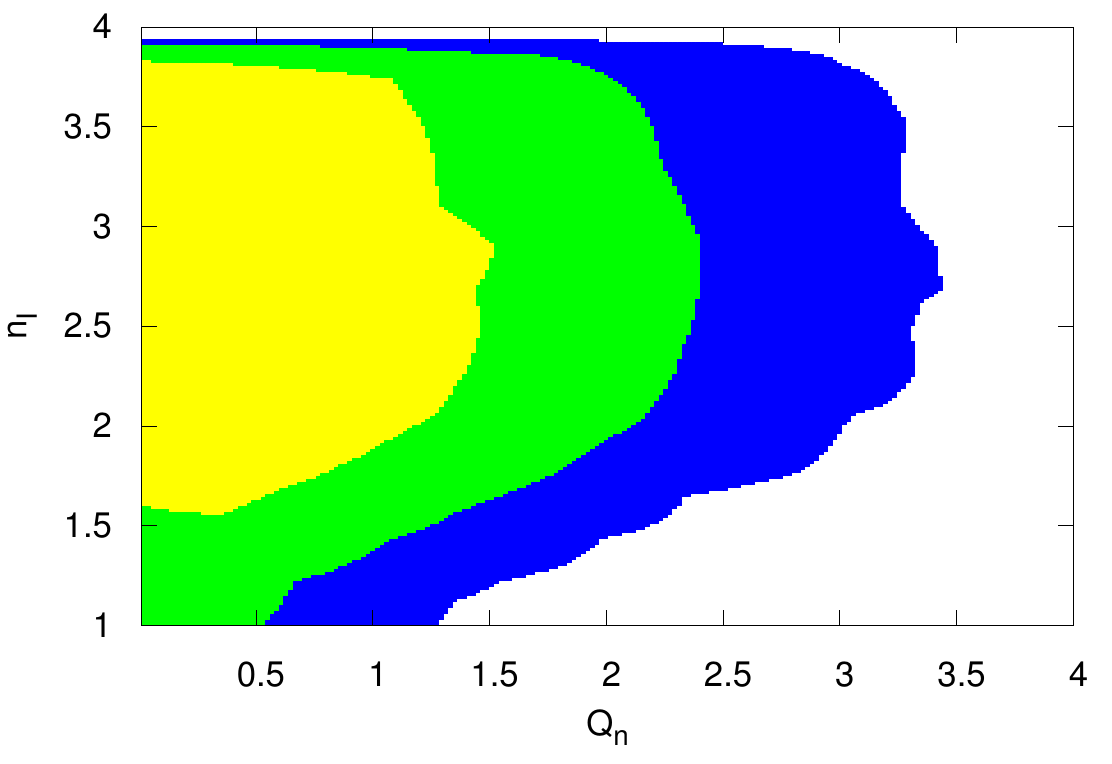} \includegraphics[width=2.1in]{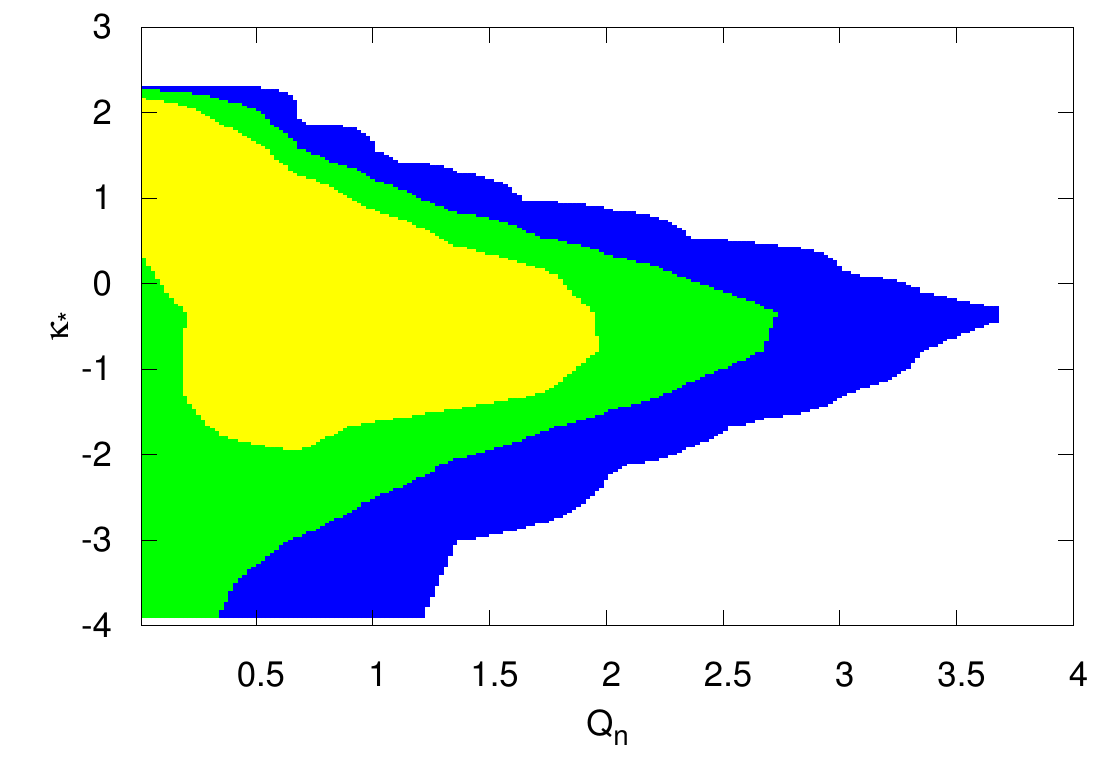}
\includegraphics[width=2.1in]{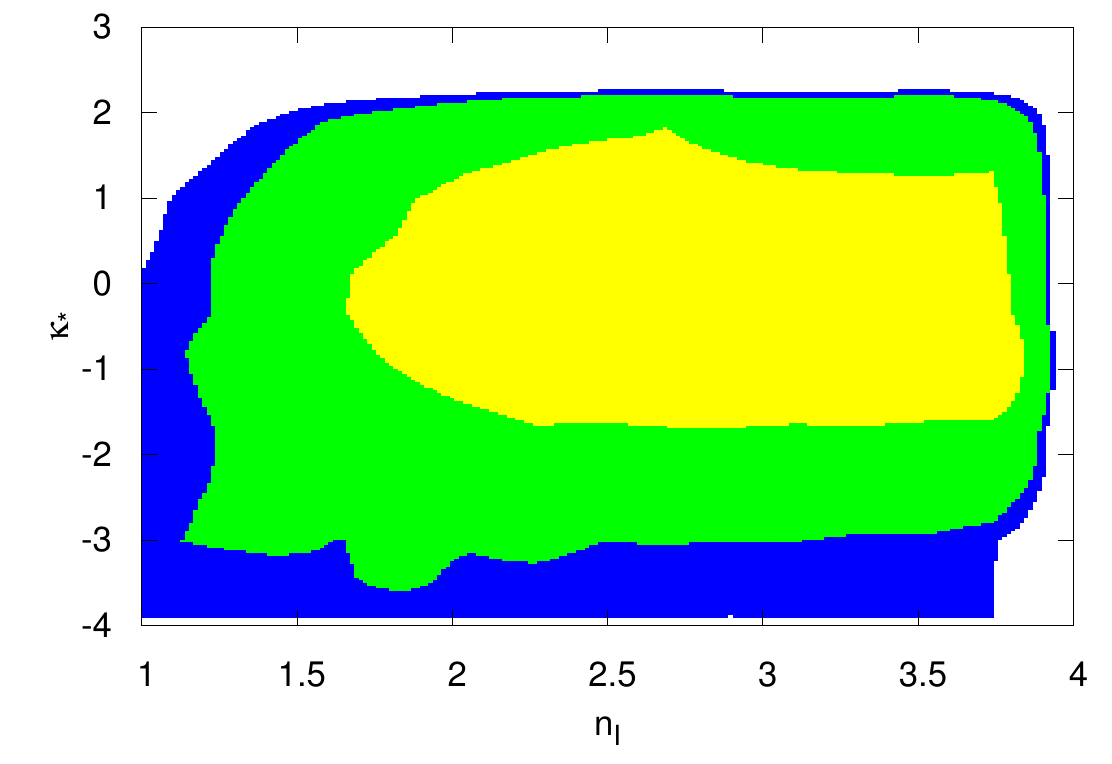}
\par\end{centering}

\centering{}\protect\caption{ Constraints on the KK model using Planck and BOSS DR11 data. Light
(yellow), medium (green), and dark (blue) shaded regions identify
$68\%$, $95\%$, and $99.7\%$ confidence contours, respectively.
\label{f:KK_PB_iso_constraints} }
\end{figure}

Comparing Planck+BOSS isocurvature constraints (e.g.~the third column
of Table~\ref{t:constraints1d_cmb+gal}) to the Planck-only constraints
of Table~\ref{t:constraints1d_cmb}, we see that $\kappastar$ increases
by $\approx0.3$ with the addition of galaxy survey data, while $\ln(\sigma_{8})$
and $\tau$ both drop by $\approx0.03$ in a way that leaves $\sigma_{8}\exp(-\tau)$
nearly constant. As with the Planck-only analysis, we see that $Q_{n}=0$
is allowed at $1\sigma$ in the KK and NB models, and at somewhat
more than $1\sigma$ in the PWR model, indicating no significant preference
for these isocurvature models. Once again, nearly the entire range
of $n_{I}$ and $\kappastar$ are within the $95\%$ confidence regions.
Comparing the two-dimensional constraints in Fig.~\ref{f:KK_PB_iso_constraints}
to those in Fig.~\ref{f:KK_P_iso_constraints}, we see slight hints
of a preference for higher $\kappastar$, $n_{I}$, and $Q_{n}$ when
galaxy survey data are included.

{\renewcommand{\baselinestretch}{1.0}   \input{table4_lamppost.tex}    } 

\begin{figure}[tb]
\begin{centering}
\includegraphics[width=2.125in]{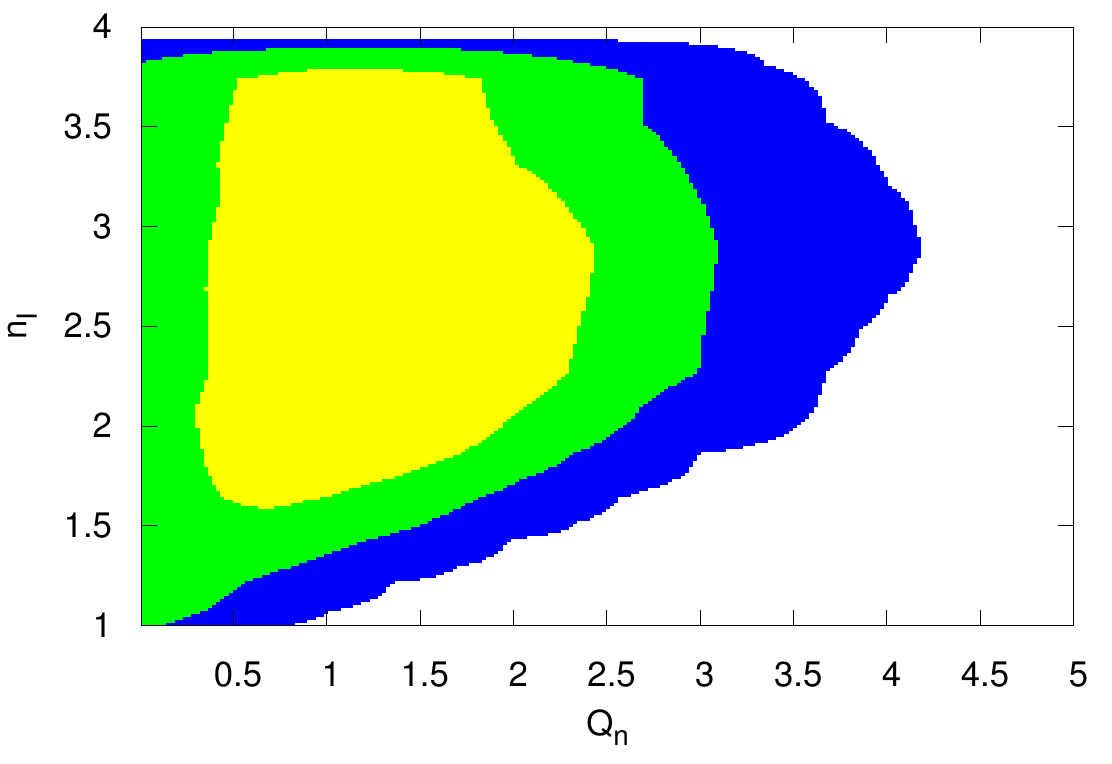} \includegraphics[width=2.125in]{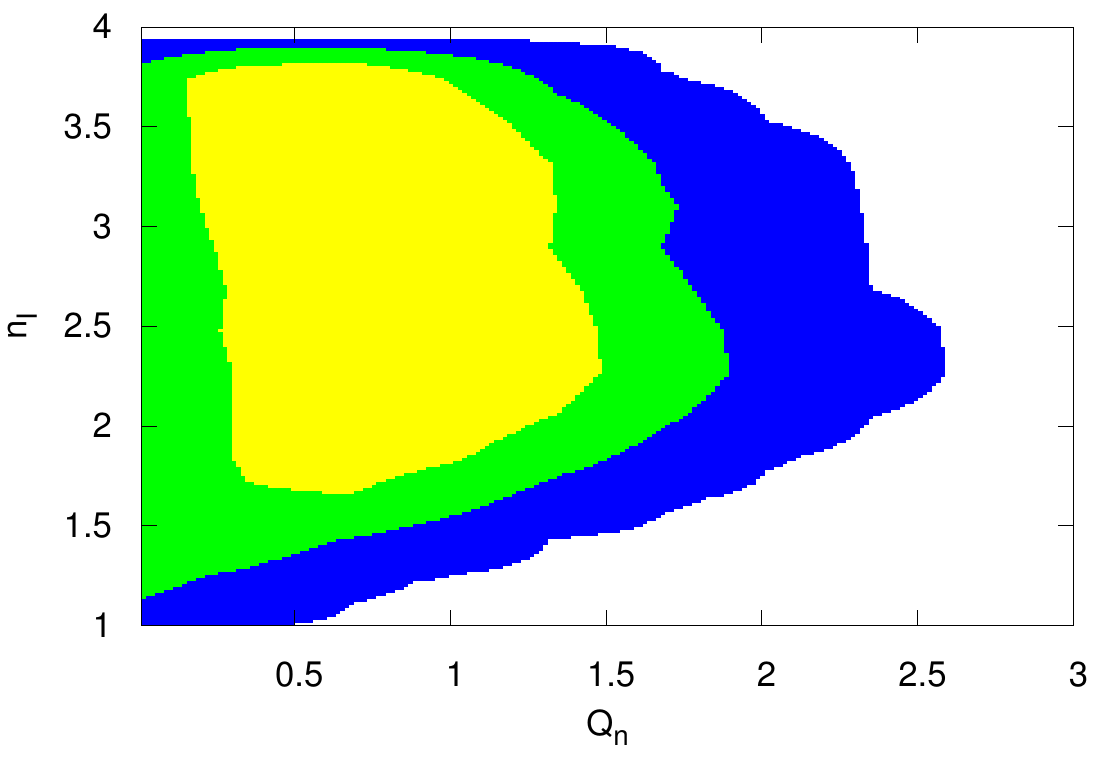}
\includegraphics[width=2.125in]{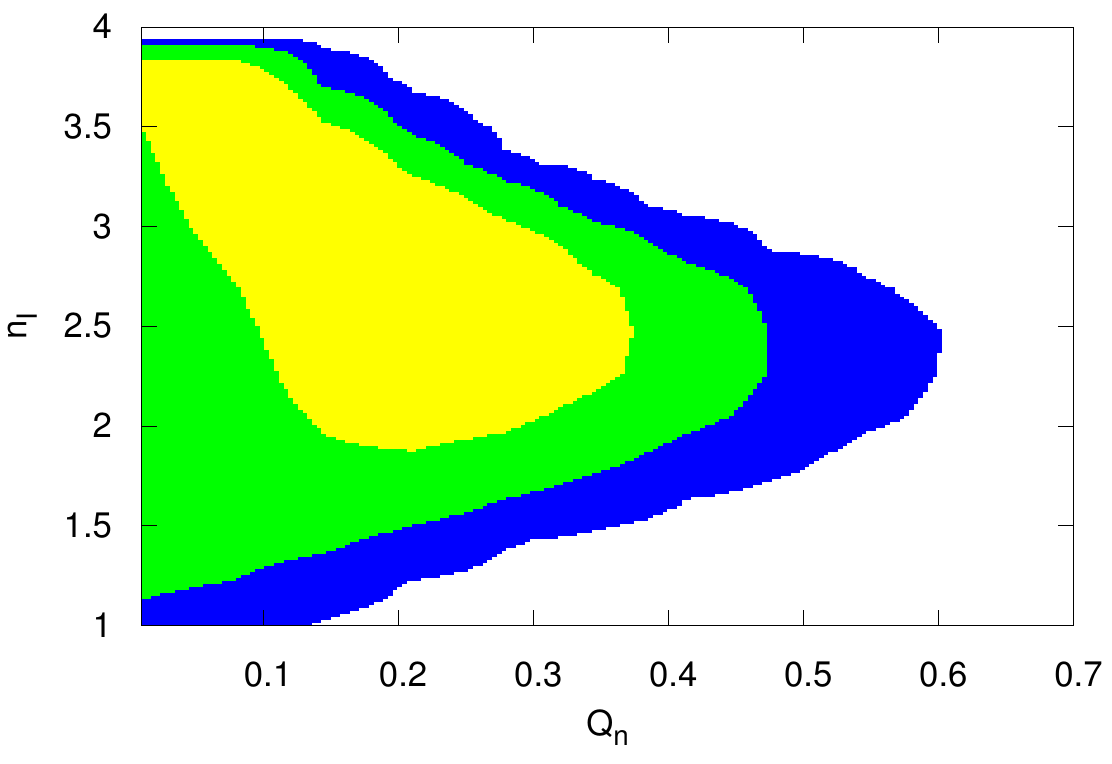}
\par\end{centering}

\centering{}\protect\caption{ Constraints on the LAMP-$1$ (left), LAMP-$2$ (middle), and LAMP-$10$
(right) models using Planck and BOSS DR11 data. Light (yellow), medium
(green), and dark (blue) shaded regions identify $68\%$, $95\%$,
and $99.7\%$ confidence contours, respectively. \label{f:LAMP_PB_iso_constraints} }
\end{figure}

KK-type isocurvature models with different $\kappastar$ are qualitatively
very different. In the small-$\kappastar$ limit, the KK model reduces
to a flat isocurvature, with weak constraints on $n_{I}$ coming only
from cosmic-variance-limited measurements at horizon scales. Thus
we consider a few specific lamppost models in which $\kappastar$
is fixed to larger values, in which current data can probe the blue-tilted
region of the power spectrum. Table~\ref{t:constraints1d_lamppost}
and Figure~\ref{f:LAMP_PB_iso_constraints} show the resulting constraints
where we have chosen the maximum $\kappastar$ to be 2.3 partly based
on the fact it is $2\sigma$ allowed by the third column of Table~\ref{t:constraints1d_cmb+gal}
(and this corresponds to the maximum of the range allowed in the MCMC
sampling as noted before). In all three cases considered, with $\kappastar\geq0$,
we see a $>1\sigma$ preference for $Q_{n}>0$.

{\renewcommand{\baselinestretch}{1.0}   \input{table5_blue_hiblue.tex}   } 

\begin{figure}[tb]
\centering{}\includegraphics[width=4in]{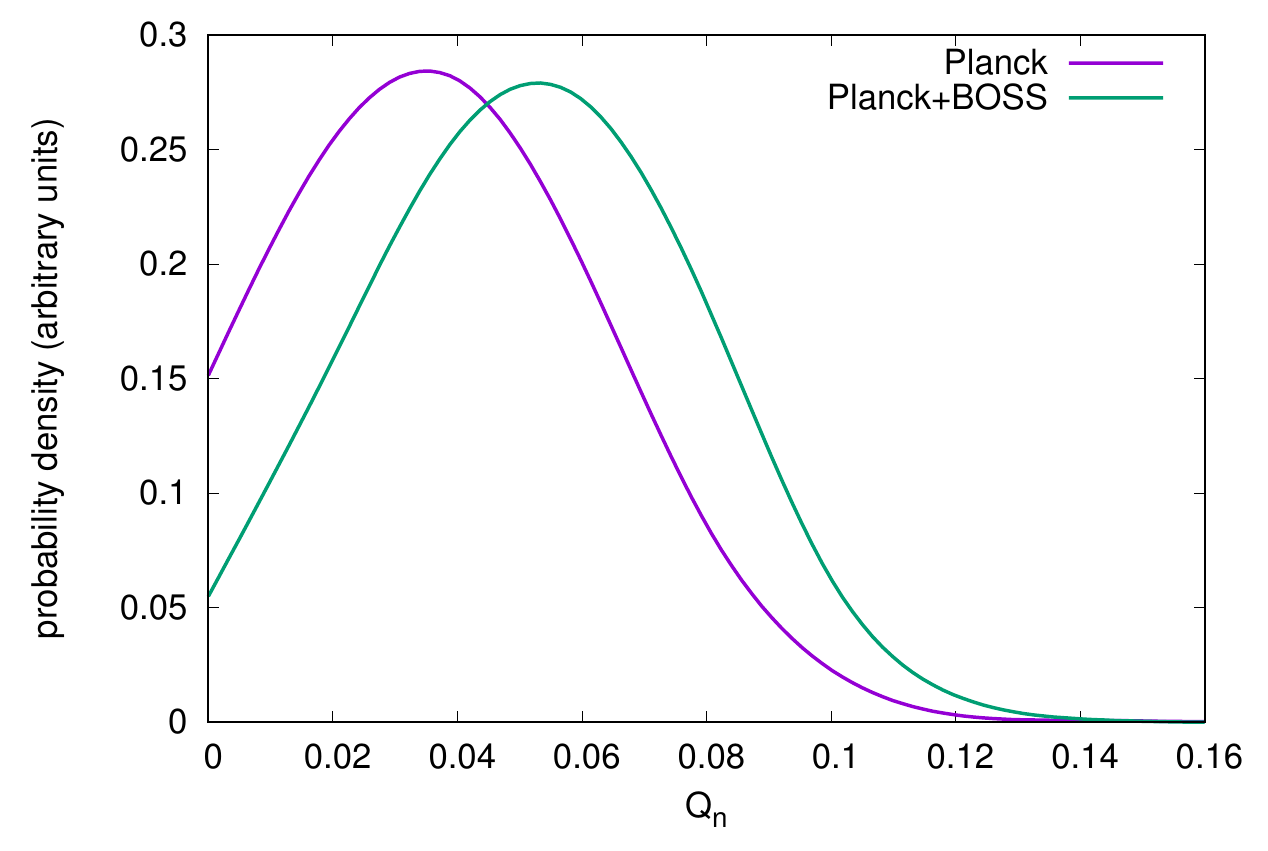}\protect\caption{ Marginalized probability density of $Q_{n}$ in the HI-BLUE model,
constrained using either Planck or Planck+BOSS data. The left (right)
curve corresponds to the Planck only (Planck+BOSS) fit.\label{f:constraints1d_HI-BLUE} }
\end{figure}

Finally, since we are specifically interested the bluest-tilted models,
we consider lamppost models in which $n_{I}=3.9$ is fixed (this value
is $2\sigma$ allowed by the third column of Table~\ref{t:constraints1d_cmb+gal}
and lies at the maximum of the allowed MCMC sampling). Constraints
on BLUE (variable-$\kappastar$) and HI-BLUE ($\kappastar=\ln(10)$)
models are shown in the final two columns of Table~\ref{t:constraints1d_lamppost}.
Intriguingly, the HI-BLUE model has a $2\sigma$ preference for $Q_{n}>0$.
We investigate this further in Table~\ref{t:constraints1d_blue_hiblue},
showing constraints with and without BOSS data. The corresponding
one-dimensional probability density is shown in Fig.~\ref{f:constraints1d_HI-BLUE}.
Even though this is encouraging, this does not represent a statistically
significant hint since there is no a priori reason to prefer the HI-BLUE
spectrum for the fits.

Since $\kappastar=\ln(10)$ corresponds to $k_{\star}/a_{0}=0.5/$~Mpc~$\approx0.7~h/$Mpc,
a few times larger than currently-accessible scales, this constraint
can be interpreted as a $2\sigma$ preference for a highly blue-tilted
isocurvature with a power-law spectrum on observable scales if there
is some reason to expect $n_{I}=3.9$ a priori.  
While a $2\sigma$ hint is hardly conclusive and there is no compelling
reason to expect $n_{I}=3.9$, if we interpret this really as a hint,
there is some reason to be optimistic about its case being strengthened
by data in the near future. Planned CMB and large-scale structure
surveys promise more sensitivity over a larger range of scales. Surveys
mapping the neutral hydrogen in the universe using the $21$~cm line
are expected to reach $k/a_{0}\sim10~h/$Mpc in the coming decades.
Such probes will shed light on physics at the highest energies through
their sensitivity to ABI models. 

Regardless of this fit result being interpreted as a hint, note that
this class of models also ``predicts'' $k_{\star}$, the break point
in the spectrum, to be in the observable range if one restricts the
theoretical bias to having sub-Planckian scalar field values and more
importantly the total number of efolds of inflation not being smaller
than around 50. More specifically, one can see from generalizing the
model dependent Eq.~(\ref{eq:phipscalefact}) that 
\begin{equation}
\frac{k_{\star}}{a_{0}}\sim\left(\frac{\varphi_{{\rm init}}}{0.3M_{p}}\right)^{\frac{2}{3}}e^{-\left(N_{e}-50\right)}\left(\frac{T_{{\rm rh}}/H}{10^{-1}}\right)^{1/3}\left(\frac{H/\varphi_{{\rm fin}}}{10^{-3}}\right)^{2/3}(10\,{\rm Mpc}^{-1})\label{eq:kstarbound}
\end{equation}
where $M_{p}$ is the reduced Planck mass, $H$ is the expansion rate
during inflation, and $\varphi$ is a model dependent order parameter
that controls whether the isocurvature perturbation modes are massive
or massless (when compared to $H$) at the time of mode horizon exit.
In an axion model specific to Eq.~(\ref{eq:phipscalefact}), $\varphi$
has the order of magnitude of the PQ symmetry breaking field $|\Phi_{+}|$.
The variable $\varphi_{{\rm init}}$ is the $\varphi$ value at $N_{e}$
efolds before the end of inflation, and the variable $\varphi_{{\rm fin}}$
is the $\varphi$ value at the time when the modes are first massless
at horizon exit.

Another positive indication for future observability of this class
of models can be seen as follows. According to column 4 of Table~\ref{t:constraints1d_blue_hiblue},
the 95\% confidence level upper bound on $Q_{n}$ is 0.11, which corresponds
to $\mathcal{Q}_{1}\approx9\times10^{-8}$. This implies that the
isocurvature power at $k/a_{0}\gtrsim0.5\,\,{\rm Mpc}^{-1}$ primordially
can be 40 times larger than the adiabatic power (in contrast with
the percent level power of a scale-invariant spectrum). Moreover,
because the data set used here is already insensitive to the spectrum
at this large $k/a_{0}\gtrsim k_{\star}/a_{0}\approx0.5\,\,{\rm Mpc}^{-1}$,
it is possible to dramatically further increase the isocurvature power
relative to the adiabatic power by increasing $k_{\star}/a_{0}$.

\section{\label{sec:Lamp-post-model-interpretation}A model interpretation}

In this section, we interpret the fit results of the last section
in terms of the axion model of \cite{Kasuya:2009up}. We will find
that the fit is consistent with a very plausible supersymmetric QCD
axion model. In particular, we will find that the result is consistent
with a scenario in which all of the dark matter is composed of axions
and the initial misalignment angle is of order unity.

The supersymmetric model \cite{Kasuya:2009up} has its axion residing
in a linear combination of PQ-charged beyond-the-Standard-Model fields
$\Phi_{+}$ and $\Phi_{-}$ where the subscripts refer to the PQ charges.
As explained in \cite{Kasuya:2009up} (and \cite{Chung:2015pga}),
the relevant effective potential during inflation is 
\begin{equation}
V\approx h_{1}^{2}|\Phi_{+}\Phi_{-}-F_{a}^{2}|^{2}+c_{+}H^{2}|\Phi_{+}|^{2}+c_{-}H^{2}|\Phi_{-}|^{2}
\end{equation}
where $\{h_{1},\, c_{\pm},\, F_{a},\, H\}$ are numerical constants.
The variable $H$ has the interpretation of the expansion rate during
inflation, and $F_{a}$ is related to the usually quoted axion decay
constant $f_{a}$ through 
\begin{equation}
f_{a}=\sqrt{2}\left(|\Phi_{+}(t_{f})|^{2}+|\Phi_{-}(t_{f})|^{2}\right)^{1/2}
\end{equation}
where 
\begin{equation}
|\Phi_{\pm}(t_{f})|=F_{a}\left(\frac{c_{\mp}}{c_{\pm}}\right)^{1/4}\sqrt{1-\frac{\sqrt{c_{+}c_{-}}H^{2}}{h_{1}^{2}F_{a}^{2}}}.
\end{equation}
Because of the insensitivity of the ABI spectrum with $h_{1}$-variation
in the parameter region of interest, we can set $h_{1}=1$ as long
as $h_{1}F_{a}\gg H$. The initial condition for $\Phi_{\pm}$ is
parameterized by 
\begin{equation}
\Phi_{\pm}(t_{i})=|\Phi_{\pm}(t_{i})|e^{\mp i\theta_{+}(t_{i})}
\end{equation}
where $\theta_{+}(t_{i})\sim O(0.1)$ for ``natural'' scenarios.

The key initial condition is that $\{|\Phi_{\pm}(t_{i})|\gg F_{a},\,\,\,\Phi_{+}(t_{i})\Phi_{-}(t_{i})\approx F_{a}^{2}\}$
and $\Phi_{+}$ rolls towards the minimum during inflation. With the
parameterization

\begin{equation}
\Phi_{\pm}\equiv\frac{\varphi_{\pm}}{\sqrt{2}}\exp\left(i\frac{a_{\pm}}{\sqrt{2}\varphi_{\pm}}\right)\label{eq:angularparam-1}
\end{equation}
where $\varphi_{\pm}$ and $a_{\pm}$ are real, the axion is 
\begin{equation}
a=\frac{\varphi_{+}}{\sqrt{\varphi_{+}^{2}+\varphi_{-}^{2}}}a_{+}-\frac{\varphi_{-}}{\sqrt{\varphi_{+}^{2}+\varphi_{-}^{2}}}a_{-}
\end{equation}
and this field will have a mass-squared that is approximately $c_{+}H^{2}$
during inflation if $|\Phi_{+}|\gg F_{a}$ while $\Phi_{+}\Phi_{-}=F_{a}^{2}$.
The Goldstone theorem is evaded because the radial field $\Phi_{+}$
is rolling and not at its minimum. This temporary massive behavior
of the axion is responsible for the blueness of the ABI spectrum.
The approximate constant behavior of the mass until $\Phi_{+}$ reaches
$\Phi_{+}(t_{f})$ is natural within supersymmetric models since the
leading SUSY breaking is controlled through gravity mediated contribution
$H$, the expansion rate, which is approximately constant during inflation.

As explained in \cite{Chung:2016wvv}, the parameter $\mathcal{Q}_{1}$
(related to the more practical fit parameter $Q_{n}$ through Eq.~(\ref{eq:Q1QNrelation}))
fixed through the fit constrains underlying model parameters through
\begin{equation}
\mathcal{Q}_{1}=\left(\frac{H}{2\pi}\right)^{2}\frac{\tilde{A}(c_{+})\sqrt{c_{-}/c_{+}}}{F_{a}^{2}\theta_{+}^{2}(t_{i})(1+c_{-}/c_{+})}\omega_{\mathrm{a}}^{2}\label{eq:Q1eq}
\end{equation}
where $\omega_{a}\equiv\Omega_{a}/\Omega_{c}$ is the dark matter
fraction in axions and is approximately 
\begin{equation}
\omega_{\mathrm{a}}\approx W_{a}\theta_{+}^{2}(t_{i})\left(\frac{\sqrt{2}F_{a}\sqrt{\frac{c_{-}+c_{+}}{\sqrt{c_{-}c_{+}}}}}{10^{12}{\rm GeV}}\right)^{n_{PT}}.
\end{equation}
Here 
\begin{equation}
W_{a}\approx1.5\,\,\,\,\,\,\,\,\,\,\,\,\,\,\, n_{PT}\approx1.19\label{eq:ptparams}
\end{equation}
are QCD phase transition physics related parameters \cite{Kawasaki:2013ae},
and we have assumed that $c_{\pm}>0$. The $\omega_{a}$ parametric
dependence assumes that the axion relic density is dominated by the
coherent oscillations after the chiral phase transition. It also assumes
that the coherent oscillations begin when $T\gtrsim0.1$ GeV such
that the axion mass has the usual nontrivial temperature dependence
of $m_{a}\propto(\Lambda_{{\rm QCD}}/T)^{3.34}$ (see equation 9 of
\cite{Kawasaki:2013ae}). In terms of $F_{a}$, we are assuming $F_{a}\lesssim10^{17}$
GeV. For larger $F_{a}$, the relic abundance formula needs modifications,
but for this section, this will not be of interest to us because this
parameter region is not phenomenologically viable. Although one can
compute $\tilde{A}$ in terms of the interpolating function of \cite{Chung:2016wvv},
its range is 
\begin{equation}
\tilde{A}(c_{+})\approx0.92\pm0.03\label{eq:Atilparam}
\end{equation}
which means one can obtain a good approximation without computing
this accurately.

Combining Eqs.~(\ref{eq:Q1QNrelation}) and (\ref{eq:Q1eq}), we
can write all the fit parameters on the right hand side of the equation
\begin{equation}
\frac{\left[H\theta_{+}(t_{i})\right]^{2}F_{a}^{2n_{PT}-2}}{\left(10^{12}{\rm GeV}\right)^{2n_{PT}}}=\frac{(2\pi)^{2}(1+c_{-}/c_{+}(n_{I}))^{1-n_{PT}}10^{-10}\left(1+e^{n_{I}\kappa_{\star}}\right)Q_{n}}{2^{n_{PT}}\,\,\tilde{A}(c_{+})\left(\sqrt{c_{-}/c_{+}(n_{I})}\right)^{1-n_{PT}}W_{a}^{2}}\label{eq:onerelationship}
\end{equation}
where Eqs.~(\ref{eq:cplusfunc}), (\ref{eq:ptparams}), (\ref{eq:Atilparam})
and the parametric choice $\{c_{-}=0.9,\, n_{I}>1.68\}$ can be used
to complete the specification of the right hand side.%
\footnote{See the discussion below Eq.~(\ref{eq:fitfunc}) and the discussion
in \cite{Chung:2016wvv} for more information about the $c_{-}$ parametric
choice.%
} For every right hand side of Eq.~(\ref{eq:onerelationship}) specified
by the fit, this equation allows us to have an area of solutions in
$(H,\theta_{+}(t_{i}),F_{a})$ space. For every point in the solution
space, there is a $\kappa_{\star}$ related constraint in the inflationary
model/initial condition parameter $(T_{rh},N_{e},|\Phi_{+}(t_{i})|)$
space through the following equation which relates observable length
scales to these inflationary parameters: 
\begin{equation}
\left(\frac{|\Phi_{+}(t_{i})|}{F_{a}}\right)^{\frac{1}{\gamma}}e^{-\left(N_{e}-54\right)}\left(\frac{T_{{\rm rh}}}{10^{7}{\rm GeV}}\right)^{1/3}\left(\frac{H}{7\times10^{8}{\rm GeV}}\right)^{1/3}=\frac{e^{\kappa_{\star}}}{2\times10^{-4}}\left(\frac{c_{+}(n_{I})}{c_{-}}\right)^{-\frac{1}{4\gamma(n_{I})}}\label{eq:phipscalefact}
\end{equation}
where 
\begin{equation}
\gamma(n_{I})=\frac{3}{2}\left(1-\sqrt{1-\frac{4}{9}c_{+}(n_{I})}\right).
\end{equation}
Here, $T_{rh}$ is the reheating temperature (temperature at which
the universe becomes radiation dominated after inflation), $N_{e}$
is the number of e-folds between an initial time $t_{i}$ and the
end of inflation, and $g_{*S}(t_{0})$ is the effective number of
entropy degrees of freedom today. Because of the exponential on the
left hand side of Eq.~(\ref{eq:phipscalefact}), the exponential
variations in $\kappa_{\star}$ can easily be accommodated in variations
in $N_{e}$. As we will see more explicitly shortly, this means that
the break in the spectrum can be placed almost anywhere in the observable
length scales as long as the number of efolds of inflation is not
strongly constrained. For an example of assumptions that can lead
to constraints, see the discussion around Eq.~(\ref{eq:kstarbound}).

\begin{figure}
\begin{centering}
a)\includegraphics[scale=0.6]{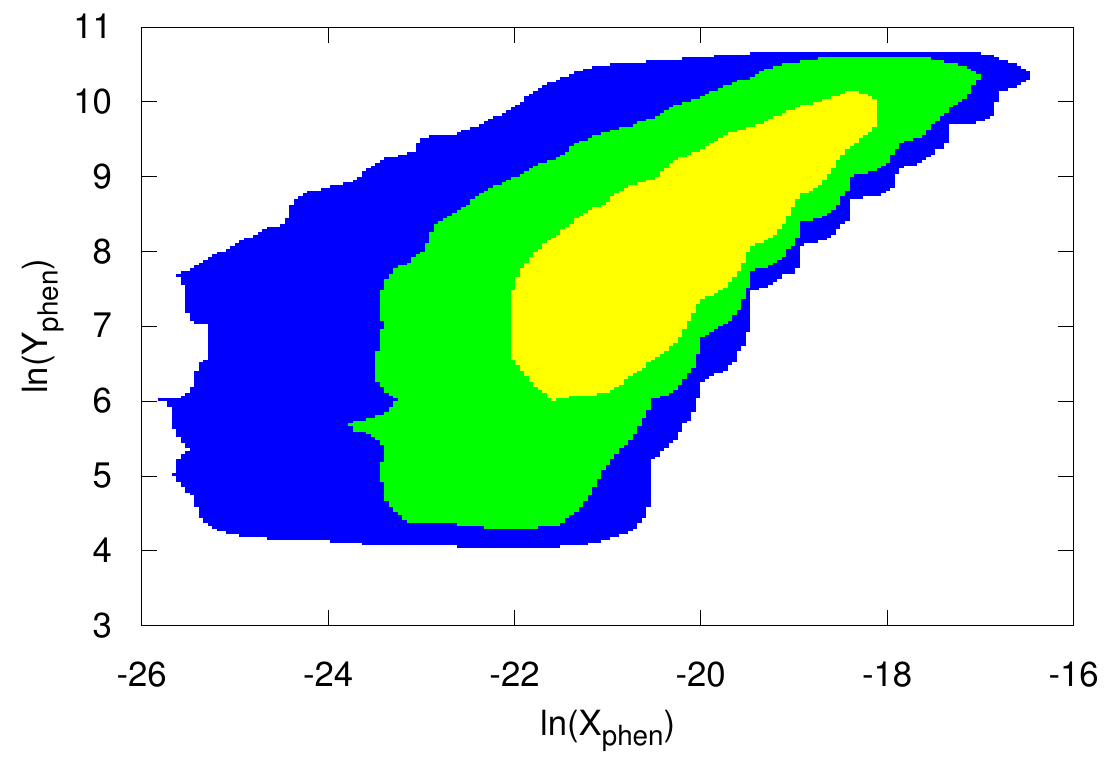}b)\includegraphics[scale=0.6]{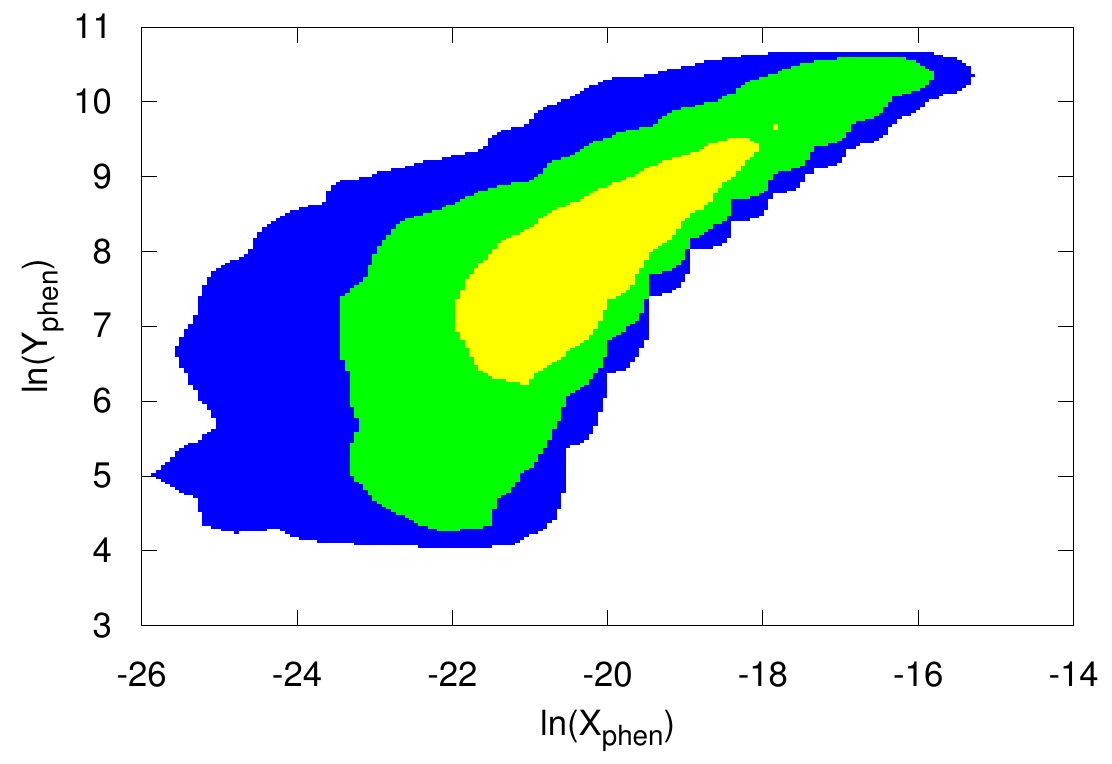} 
\par\end{centering}

\begin{centering}
c)\includegraphics[scale=0.6]{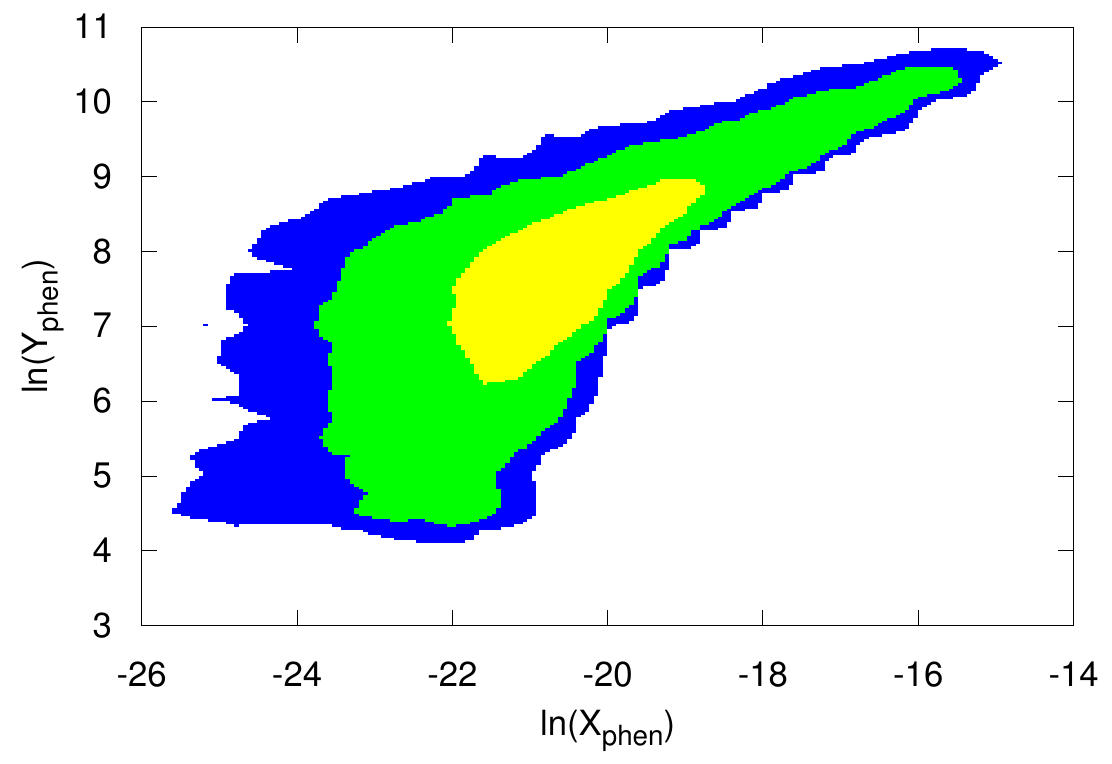} d)\includegraphics[scale=0.6]{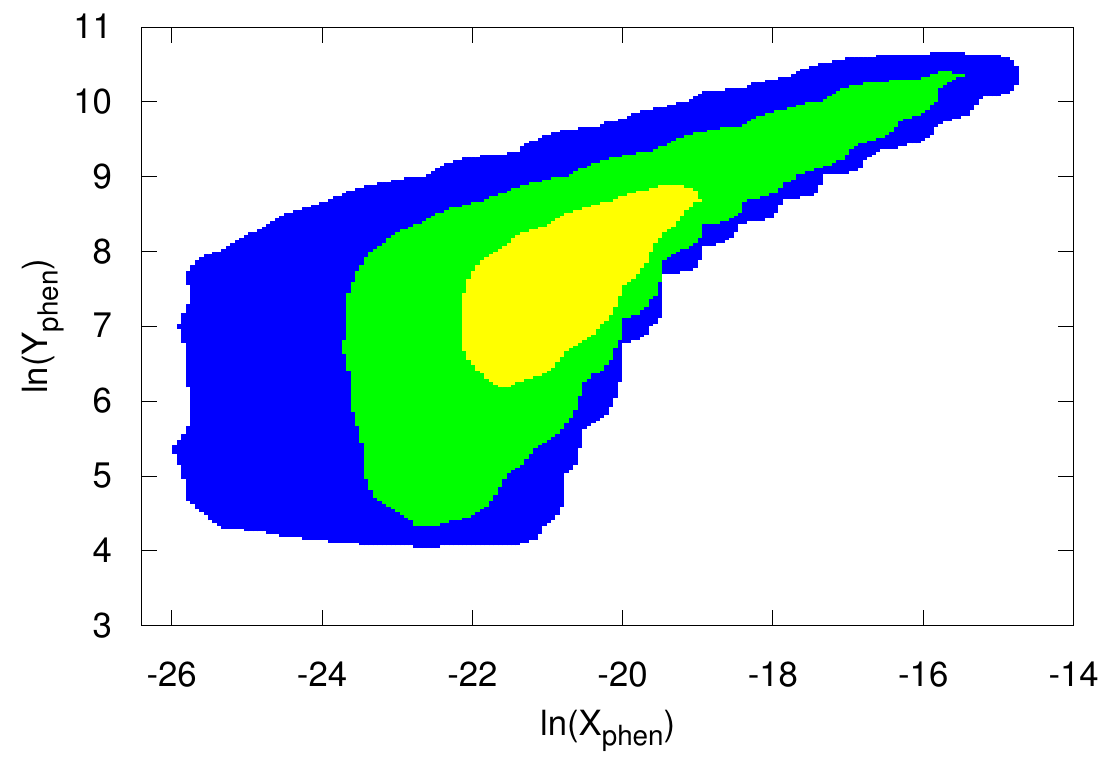} 
\par\end{centering}

\protect\protect\caption{\label{fig:xy-Distribution}Distribution of $(X_{{\rm phen}},Y_{{\rm phen}})$
(with CMB only KK fit) is plotted for (a) $n_{I}\in[1.9,2.3]$, (b)
$n_{I}\in[2.6,2.9]$, and (c) $n_{I}\in[3.54,3.94]$ (d) BLUE model
with $n_{I}=3.9$. Each successive contoured regions corresponds to
$1,2,3$-$\sigma$ regions. Since the distributions (c) and (d) are
similar, the constraint is not very sensitive to the isocurvature
spectral index $n_{I}$ ``far'' above the central spectral index
of Eq.~(\ref{eq:best}). This suggests that much of the constraint
for the ``large'' $n_{I}$ models with the current data is coming
from the bump region and above in $k$ space since that part of the
data is not as sensitive to the spectral index for $\kappa_{\star}<0$.}
\end{figure}

\begin{figure}
\centering{}\includegraphics[scale=0.75]{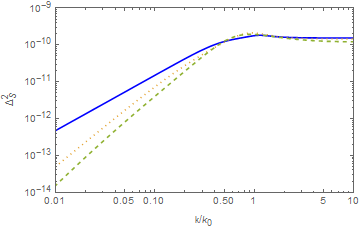}\protect\protect\caption{\label{fig:-insensitivity}$\Delta_{S}^{2}(k,k_{\star},n_{I},\mathcal{Q}_{1})$
is plotted for $n_{I}=2.5$ (solid), $n_{I}=3.2$ (dotted), and $n_{I}=3.9$
(dashed) with $k_{\star}=0.81k_{0}$ and $Q_{n}=0.96$. The bump region
and above are not very sensitive to the spectral index for a fixed
$Q_{n}$. The plateau amplitude (corresponding to the best fit) is
approximately 10\% of the adiabatic power, which represents an order
of magnitude enhancement compared to the current bounds on the flat
spectral index case.}
\end{figure}

Recall that the best fit spectral index is 
\begin{equation}
n_{I}=2.8{}_{-0.6}^{+1.1}\,\,\,\,(1{\rm \sigma})\label{eq:best}
\end{equation}
taken from the third column of Table \ref{t:constraints1d_cmb+gal}.
Given that the right hand side of Eqs.~(\ref{eq:onerelationship})
and (\ref{eq:phipscalefact}) only contain fit parameters, we plot
in Fig.~\ref{fig:xy-Distribution} the $(X_{{\rm phen}},Y_{{\rm phen}})$
distribution generated by MCMC for bins of $n_{I}$ surrounding the
best fit spectral index of Eq.~(\ref{eq:best}) where 
\begin{equation}
X_{{\rm phen}}\equiv\frac{(2\pi)^{2}(1+c_{-}/c_{+}(n_{I}))^{1-n_{PT}}10^{-10}\left(1+e^{n_{I}\kappa_{\star}}\right)Q_{n}}{2^{n_{PT}}\,\,\tilde{A}(c_{+})\left(\sqrt{c_{-}/c_{+}(n_{I})}\right)^{1-n_{PT}}W_{a}^{2}}
\end{equation}
\begin{equation}
Y_{{\rm phen}}\equiv\frac{e^{\kappa_{\star}}}{2\times10^{-4}}\left(\frac{c_{+}(n_{I})}{c_{-}}\right)^{-\frac{1}{4\gamma(n_{I})}}
\end{equation}
 and the $Q_{n}$ dependence shows up only in $X_{{\rm phen}}$. As
explained in the figure captions, the results suggest that much of
the constraint for the $n_{I}\gtrsim3$ models with the current data
is coming from the bump region and above in $k$ space since that
part of the data is not as sensitive to the spectral index for $\kappa_{\star}<0$.
The insensitivity of the $k\gtrsim k_{0}\exp(\kappa_{\star})$ part
of the spectrum with the spectral index $n_{I}$ is illustrated in
Fig.~\ref{fig:-insensitivity}. On the other hand, the likelihood
for the $n_{I}<2.8$ region is more sensitive to the data with $k$
smaller than the break (and hence the likelihood is more sensitive
to the spectral index) since the isocurvature amplitude there is not
as suppressed in the case of the smaller spectral index. This also
explains the asymmetry in the error bars in Eq.~(\ref{eq:best}).
Note that because CMB observables are not as sensitive to large $k$
isocurvature primordial spectrum compared to the large $k$ adiabatic
primordial spectrum, the $k<k_{\star}$ part of the spectrum in Fig.~\ref{fig:-insensitivity}
is more significant for CMB fits than it naively appears for shallow
$n_{I}$. 

\begin{figure}
\begin{centering}
\includegraphics[scale=0.53]{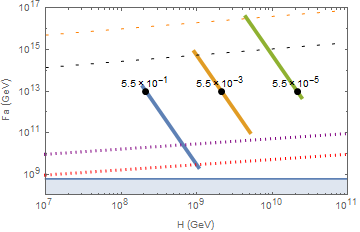}~~~~~\includegraphics[scale=0.53]{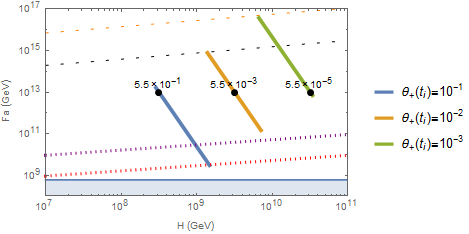} 
\par\end{centering}

\protect\protect\caption{\label{fig:Lamp-post-axionic-parameter}Shown are axionic parameter
regions consistent with the best fit parameter set $S_{1}$ (left)
and a larger $Q_{n}$ parameter set $S_{2}$ (right). Hence, the effect
of increasing $Q_{n}$ and adjusting $\kappa_{\star}$ to maintain
a good fit to the data shifts the underlying model parameters $\{H,F_{a}\}$
to the right. The upper ends on all the thick lines come from the
saturation of the dark matter bound: $\omega_{a}\leq1$. The bottom
ends on all the thick lines come from the bound of perturbativity:
$\delta\rho_{a}/\rho_{a}<1$. The dashed curves represent $|\Phi_{+}(t_{i})|<0.3M_{p}$
with two different numbers of inflationary efolds. The upper one assumes
that the number of efolds $N_{e}$ of inflation is at least $50$
while the second curve assumes that the number of efolds of inflation
is at least 54. The dotted curves towards the bottom of the figure
represent boundaries below which the total dark matter abidance may
also contain cosmic strings because the maximum temperature reached
during the reheating period is larger than $F_{a}$. The upper dotted
curve is for $T_{rh}=10^{8}$ GeV while the lower dotted curve is
for $T_{rh}=10^{6}$ GeV. The bottom blue region is excluded by the
supernova 1987A burst duration (e.g.~\cite{Raffelt:2006cw}), and
some literature exclude $F_{a}$ values that are an order of magnitude
higher \cite{Hertzberg:2008wr}. The numbers by the isolated dots
indicate $\omega_{a}$ at that point in the parameter space.}
\end{figure}

Inspired by the fit results of Fig.~\ref{fig:xy-Distribution}, Eq.~(\ref{eq:best}),
and Section \ref{sec:Data-fit}, we choose two representative parameter
sets to investigate whether interpreting these parameters in terms
of the axion model of \cite{Kasuya:2009up} leads to a reasonable
physical picture. One set we choose is the approximately best fit
set $S_{1}\equiv\{n_{I}=2.8,\,\, X_{{\rm phen}}=\exp(-20.6)\,\,,Y_{{\rm phen}}=\exp(8.1)\}$
(corresponding to $\{Q_{n}=0.96,\,\,\kappa_{\star}=-0.21,\,\mathcal{Q}_{1}=1.5\times10^{-10}\}$)
and a second set $S_{2}\equiv\{n_{I}=2.8,\,\, X_{{\rm phen}}=\exp(-19.8),\,\, Y_{{\rm phen}}=\exp(7.7)\}$
(corresponding to $\{Q_{n}=2.8,\,\kappa_{\star}=-0.6,\,\mathcal{Q}_{1}=3.3\times10^{-10}\}$)
which gives a larger $Q_{n}$ that is still $1\sigma$ consistent
with the central value in the binned distribution of Fig.~\ref{fig:xy-Distribution}b).
The $\{\theta_{+}(t_{i}),H,F_{a}\}$ parameter regions consistent
with $S_{1}$ and $S_{2}$ are shown in Fig.~\ref{fig:Lamp-post-axionic-parameter}.
The most important phenomenological self-consistency constraint in
Fig.~\ref{fig:Lamp-post-axionic-parameter} is that the axion dark
matter does not exceed the totality of cold dark matter abundance:
\begin{equation}
\omega_{a}\leq1.\label{eq:overclosure}
\end{equation}
This determines the upper ends of each of the allowed $(H,F_{a})$
curves. The most important theoretical constraint comes from the validity
of the linear computation 
\begin{equation}
\frac{\sqrt{\Delta_{S}^{2}}}{\omega_{a}}<1
\end{equation}
which is a restatement of the assumed smallness of axion energy overdensity
$\delta\rho_{a}/\rho_{a}<1$. Since the spectral peak is less than
about twice the plateau, we can impose a simpler bound 
\begin{equation}
\frac{\mathcal{Q}_{1}}{\omega_{a}}<\frac{1}{2}
\end{equation}
which will set a lower bound on $F_{a}$. This determines the lower
ends of each of the allowed $(H,F_{a})$ curves.

The dashed curves in Fig.~\ref{fig:Lamp-post-axionic-parameter}
represent $|\Phi_{+}(t_{i})|<0.3M_{p}$ with two different number
of inflationary efolds. If $|\Phi_{+}(t_{i})|$ is above this value,
we would generically be wary of the breakdown of the effective field
theory description that neglects gravity suppressed non-renormalizable
operators. Note that $N_{e}$ in Eq.~(\ref{eq:phipscalefact}) represents
the number of efolds between time $t_{i}$ and the end of inflation.
Hence, we see from the figure that the initial non-equilibrium value
of $|\Phi_{+}(t_{i})|$ need not be very large to satisfy the best
fit value of $\kappa_{\star}$. The dotted curves towards the bottom
of the plot represent the boundary below which we would have to take
into account the cosmic string decay contribution to the axionic dark
matter abundance due to the fact that PQ symmetry might be restored
if 
\begin{equation}
F_{a}\lesssim T_{{\rm max}}=(0.77)\left(\frac{9}{5\pi^{3}g_{*}}\right)^{1/8}\sqrt{T_{rh}}\left(HM_{p}\sqrt{8\pi}\right)^{1/4}
\end{equation}
where $T_{{\rm max}}$ is taken from \cite{Chung:1998rq} and we have
assumed in Fig.~\ref{fig:Lamp-post-axionic-parameter} that the number
of degrees of freedom $g_{*}$ contributing to the energy density
is 200 at the completion of reheating. If the axionic string network
reaches scaling regime, then the decay of the strings will contribute
an axion abundance of \cite{Kawasaki:2013ae} 
\begin{equation}
\Omega_{a,{\rm str}}\approx2.0\xi\left(\frac{F_{a}}{10^{12}{\rm GeV}}\right)^{1.19}\left(\frac{\Lambda_{QCD}}{400{\rm MeV}}\right)\label{eq:additionalstring}
\end{equation}
which would be relevant in the parameter regime below the dotted curve
in Fig.~\ref{fig:Lamp-post-axionic-parameter}. Since there is a
large parameter region in which axions constitute all of dark matter,
we will not dwell on this parametric corner where the string contribution
becomes important.

Some other constraints that we have considered but are not important
in the best fit parameter region are the following. Making sure that
the initial $\theta_{+}(t_{i})$ tuning is above the quantum noise
and noting the approximation made in equation 29 of \cite{Chung:2016wvv},
we impose 
\begin{equation}
\frac{H}{2\pi|\Phi_{+}(t_{i})|}\ll\theta_{+}(t_{i})\ll1.
\end{equation}
If we require that the classical value of the conserved quantity be
always greater than the quantum fluctuations (not just at the initial
time), we would end up with a stronger constraint 
\begin{equation}
\frac{H}{4\pi F_{a}\sqrt{c_{-}+c_{+}}}\ll\theta_{+}(t_{i})\ll1.
\end{equation}
These constraints are not as strong as the ones playing a role in
Fig.~\ref{fig:Lamp-post-axionic-parameter}.

It is important to note that the axionic degree of freedom naturally
carries both adiabatic and isocurvature inhomogeneity condition information
because of the gravitational coupling between the inflaton and the
axion, as discussed in \cite{Chung:2015pga}. In spatially flat gauge,
this imprinting of the adiabatic inhomogeneities shows up as a secular
time integral effect. Hence, even though the axion is a spectator
field with its own independent quantum fluctuations, it naturally
acquires mixed boundary conditions.

\section{\label{sec:Conclusions}Conclusions}

In this work, we have fit the ABI spectrum to Planck and BOSS DR11
data. Unlike the usual isocurvature spectrum that is fit to data in
the literature, this spectrum has a strong blue tilt up to $k_{\star}$,
has a little bump, and is flat beyond that. We used the economical
3-parameter fitting function of \cite{Chung:2016wvv} in the context
of 6-parameter vanilla $\Lambda$-CDM and find that the data mildly
prefers this type of isocurvature contribution. In particular, we
find the best-fit isocurvature parameter set of about $\{k_{\star}/a_{0}=4.1_{-2.7}^{+14}\times10^{-2}{\rm Mpc}^{-1},\,\, n_{I}=2.76{}_{-0.59}^{+1.1},\,\, Q_{n}=0.96_{-0.93}^{+0.32}\}$
(1$\sigma$ error bars) which indicates a decent fit with the ABI
spectrum making up about 10\% of the power on short scales. Unfortunately,
it is clear that there is no statistical significance to this nonzero
isocurvature amplitude. Note that 10\% of the primordial power on
short scales is much larger than what one would expect from a scale-invariant
isocurvature spectrum. The rest of the $\Lambda$-CDM parameters can
be found in Table \ref{t:constraints1d_cmb+gal}. If we fix the spectral
index and the break point to be large ($n_{I}=3.9,\, k_{\star}/a_{0}=0.7~h/{\rm Mpc}$),
we find a $2\sigma$ preference for a non-zero ABI spectrum as indicated
by Fig.~\ref{f:constraints1d_HI-BLUE}. It is interesting to note
that the $2\sigma$ acceptable fit of this HI-BLUE model allows the
primordial isocurvature power to be 40 times the adiabatic primordial
power at $k\gtrsim k_{\star}$ scales.

Furthermore, in the context of the axion model of \cite{Kasuya:2009up},
the best fit parameter region corresponds to all of the dark matter
being made up of QCD axions with the axion decay constant of order
$10^{13}$ GeV and an expansion rate of order $10^{8}$ GeV during
inflation. This interpretation would imply no detection of inflaton
generated gravity waves (tensor perturbations) in the near future
(e.g.~in experiments such as CMB-S4 \cite{Abazajian:2016yjj}). However,
the axion masses would be within the range of detectability through
microwave cavity type of experiments \cite{Battaglieri:2017aum}.
Although all of these results are encouraging, the fit results are
statistically inconclusive.

\begin{figure}
\begin{centering}
\includegraphics{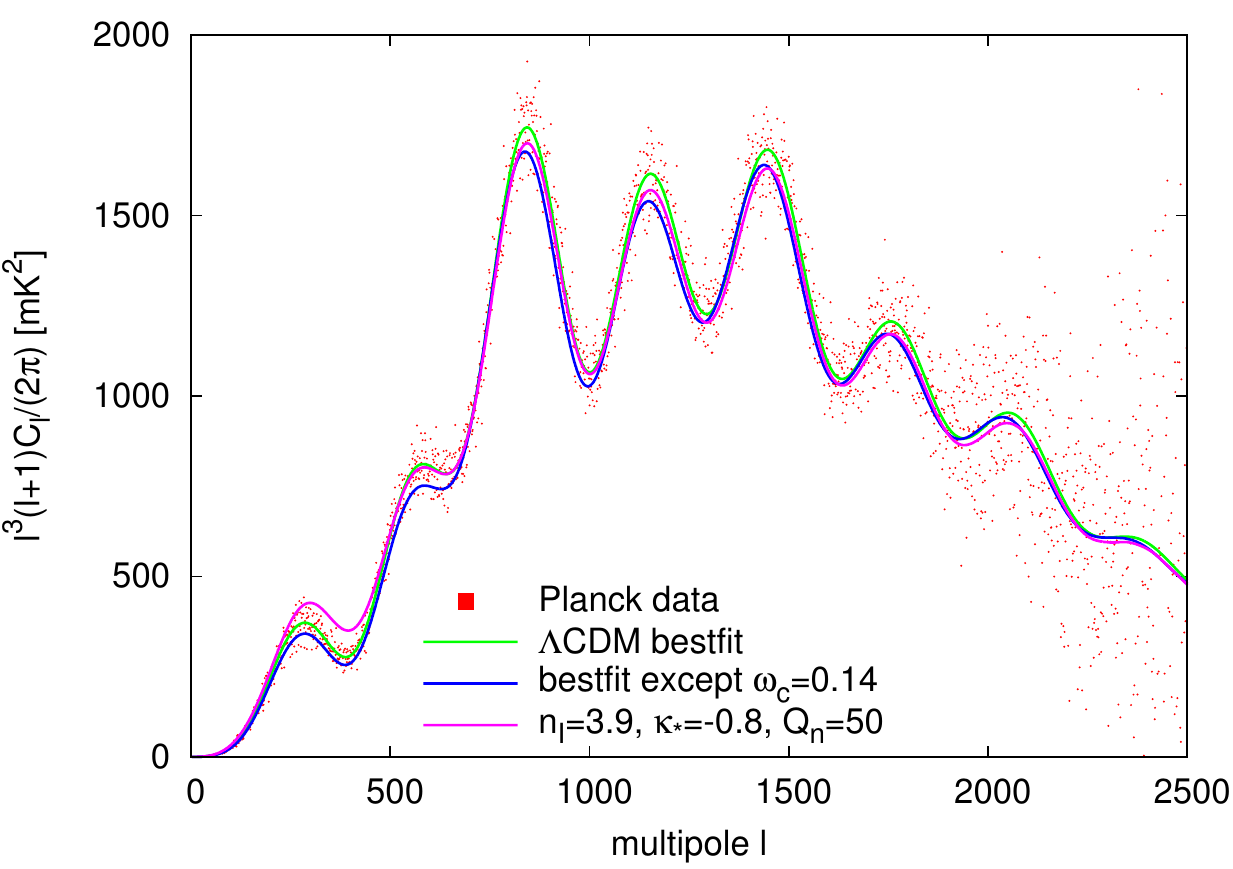} 
\par\end{centering}

\protect\protect\caption{\label{fig:Qualitative-picture-showing}Qualitative picture showing
how an exaggeratedly large $\omega_{c}\equiv\Omega_{c}h^{2}$ $\Lambda$-CDM
$C_{l}$ prediction in the high-$\ell$ region is mimicked by the
prediction with the addition of an exaggeratedly large ABI contribution.}
\end{figure}

On the other hand, there is additional reason to have some optimism
that the results are hinting at a signal. As investigated by \cite{Addison:2015wyg},
a $\Lambda$-CDM fit to small $l$ ($l\in[2,1000)$) and the large
$l$ ($l\geq1000$) Planck data gives about a 2$\sigma$ discrepant
value of $\Omega_{c}h^{2}$. In particular, the low-$\ell$ data prefers
$\Omega_{c}h^{2}\approx0.115$ while the high-$\ell$ data prefers
$\Omega_{c}h^{2}\approx0.125$. Although \cite{Addison:2015wyg} disfavors
this discrepancy as a hint for new physics because of the Planck data's
tension with the South Pole Telescope data, the interpretation of
this discrepancy is currently unresolved, and what we may be detecting
in the ABI fit presented in this paper is this mismatch between the
low-$\ell$ and high-$\ell$ data. For example, reference \cite{Addison:2015wyg}
considered the possibility of increasing a CMB lensing phenomenological
parameter $A_{L}$ (possibly motivated by modified gravity) to resolve
the anomaly. The paper \cite{Valiviita:2017fbx} shows that the $A_{L}$
can be set to its general relativity value of $A_{L}=1$ using compensated
isocurvature perturbations. One can obtain a sense of how the ABI
spectrum mimics the large $\Omega_{c}h^{2}$ effect in the large $\ell$
region through Fig.~\ref{fig:Qualitative-picture-showing}, which
significantly exaggerates both $\Omega_{c}h^{2}$ and $Q_{n}$ to
make the effect more apparent.

Although future data may shed light on the systematics between the
low-$\ell$ and the high-$\ell$, the current state of the data seems
unclear. For example, the SPTpol polarization data of \cite{Henning:2017nuy}
for $l<1000$ is consistent with high $\Omega_{c}h^{2}$ while the
data for $l>1000$ prefers a large $\Omega_{c}h^{2}$. The ACTPol
data of \cite{Louis:2016ahn} has error bars that are consistent with
both high and low $\Omega_{c}h^{2}$. Although the most probable interpretation
of the low-$\ell$ vs.~high-$\ell$ anomaly can be argued to be the
existence of yet not well understood systematics, if it is a signal
of new physics, we can look forward to future data increasing the
statistical significance of the hint. Indeed, planned CMB and large-scale
structure surveys will improve data sensitivity over a larger range
of scales. Since experiments measuring the $21$~cm line are expected
to reach scale sensitivities of $k/a_{0}\sim10~h/$Mpc in the coming
decades \cite{Villaescusa-Navarro:2014rra}, such probes may shed
light on physics at the highest energies by confirming or excluding
hints of ABI perturbations.
\begin{acknowledgments}
This work was supported in part by the DOE through grant DE-SC0017647.
\textquotedbl{}This research was performed using the compute resources
and assistance of the UW-Madison Center For High Throughput Computing
(CHTC) in the Department of Computer Sciences. The CHTC is supported
by UW-Madison, the Advanced Computing Initiative, the Wisconsin Alumni
Research Foundation, the Wisconsin Institutes for Discovery, and the
National Science Foundation, and is an active member of the Open Science
Grid, which is supported by the National Science Foundation and the
U.S.~Department of Energy's Office of Science 
\end{acknowledgments}
\bibliographystyle{JHEP}
\bibliography{Inflation_general,ref,misc,blue_isocurvature,blueiso_constraints}

\end{document}

%% file: table2_cmb.tex

\begin{table*}[tb]
  \tabcolsep=0.02cm
  \begin{footnotesize}
    \begin{tabular}{r||c|c|c|c}
      &
      KK, P(TT-only)
      &
      KK, P
      &
      NB, P
      &
      PWR, P
      \\
      
      \hline
\hline
$n_s$
&
 $0.9684 \begin{array}{ll} {}_{+0.0073} & {}_{+0.014} \\ {}^{-0.0071} & {}^{-0.014} \end{array}$
&
 $0.9658 \begin{array}{ll} {}_{+0.0056} & {}_{+0.011} \\ {}^{-0.0051} & {}^{-0.01} \end{array}$
&
 $0.9656 \begin{array}{ll} {}_{+0.0049} & {}_{+0.011} \\ {}^{-0.0054} & {}^{-0.01} \end{array}$
&
 $0.9646 \begin{array}{ll} {}_{+0.0049} & {}_{+0.009} \\ {}^{-0.0053} & {}^{-0.0092} \end{array}$
\\
\hline
$\sigma_8$
&
 $0.851 \begin{array}{ll} {}_{+0.023} & {}_{+0.043} \\ {}^{-0.022} & {}^{-0.045} \end{array}$
&
 $0.844 \begin{array}{ll} {}_{+0.019} & {}_{+0.036} \\ {}^{-0.016} & {}^{-0.038} \end{array}$
&
 $0.843 \begin{array}{ll} {}_{+0.018} & {}_{+0.035} \\ {}^{-0.017} & {}^{-0.036} \end{array}$
&
 $0.841 \begin{array}{ll} {}_{+0.021} & {}_{+0.036} \\ {}^{-0.015} & {}^{-0.042} \end{array}$
\\
\hline
$h$
&
 $0.6799 \begin{array}{ll} {}_{+0.011} & {}_{+0.022} \\ {}^{-0.011} & {}^{-0.022} \end{array}$
&
 $0.6765 \begin{array}{ll} {}_{+0.0075} & {}_{+0.014} \\ {}^{-0.0071} & {}^{-0.014} \end{array}$
&
 $0.6761 \begin{array}{ll} {}_{+0.0067} & {}_{+0.014} \\ {}^{-0.007} & {}^{-0.014} \end{array}$
&
 $0.6762 \begin{array}{ll} {}_{+0.0058} & {}_{+0.015} \\ {}^{-0.007} & {}^{-0.013} \end{array}$
\\
\hline
$\omega_\mathrm{c}$
&
 $0.1184 \begin{array}{ll} {}_{+0.0022} & {}_{+0.0046} \\ {}^{-0.0023} & {}^{-0.0046} \end{array}$
&
 $0.119 \begin{array}{ll} {}_{+0.0015} & {}_{+0.0031} \\ {}^{-0.0017} & {}^{-0.0031} \end{array}$
&
 $0.119 \begin{array}{ll} {}_{+0.0015} & {}_{+0.0031} \\ {}^{-0.0015} & {}^{-0.003} \end{array}$
&
 $0.119 \begin{array}{ll} {}_{+0.0015} & {}_{+0.0029} \\ {}^{-0.0013} & {}^{-0.0033} \end{array}$
\\
\hline
$\omega_\mathrm{b}$
&
 $0.02236 \begin{array}{ll} {}_{+0.00027} & {}_{+0.00053} \\ {}^{-0.00025} & {}^{-0.00049} \end{array}$
&
 $0.02228 \begin{array}{ll} {}_{+0.00016} & {}_{+0.00032} \\ {}^{-0.00017} & {}^{-0.00033} \end{array}$
&
 $0.02227 \begin{array}{ll} {}_{+0.00017} & {}_{+0.00032} \\ {}^{-0.00017} & {}^{-0.00033} \end{array}$
&
 $0.02228 \begin{array}{ll} {}_{+0.00018} & {}_{+0.00032} \\ {}^{-0.00016} & {}^{-0.0003} \end{array}$
\\
\hline
$\tau$
&
 $0.108 \begin{array}{ll} {}_{+0.034} & {}_{+0.06} \\ {}^{-0.028} & {}^{-0.063} \end{array}$
&
 $0.0976 \begin{array}{ll} {}_{+0.026} & {}_{+0.049} \\ {}^{-0.023} & {}^{-0.051} \end{array}$
&
 $0.0968 \begin{array}{ll} {}_{+0.025} & {}_{+0.049} \\ {}^{-0.023} & {}^{-0.049} \end{array}$
&
 $0.0904 \begin{array}{ll} {}_{+0.026} & {}_{+0.048} \\ {}^{-0.021} & {}^{-0.049} \end{array}$
\\
\hline
$Q_n$
&
 $1.2 \begin{array}{ll} {}_{+0.3} & {}_{+2.4} \\ {}^{-1.2} & {}^{-1.2} \end{array}$
&
 $1.0 \begin{array}{ll} {}_{+0.3} & {}_{+1.4} \\ {}^{-1.0} & {}^{-1.0} \end{array}$
&
 $1.1 \begin{array}{ll} {}_{+0.3} & {}_{+1.5} \\ {}^{-1.1} & {}^{-1.1} \end{array}$
&
 $0.010 \begin{array}{ll} {}_{+0.003} & {}_{+0.012} \\ {}^{-0.010} & {}^{-0.010} \end{array}$
\\
\hline
$n_I$
&
 $2.75 \begin{array}{ll} {}_{+1.19} & {}_{+1.19} \\ {}^{-0.44} & {}^{-1.3} \end{array}$
&
 $2.74 \begin{array}{ll} {}_{+1.2} & {}_{+1.2} \\ {}^{-0.66} & {}^{-1.2} \end{array}$
&
 $2.63 \begin{array}{ll} {}_{+0.78} & {}_{+1.3} \\ {}^{-0.65} & {}^{-1.2} \end{array}$
&
 $2.43 \begin{array}{ll} {}_{+0.6} & {}_{+1.2} \\ {}^{-0.53} & {}^{-1} \end{array}$
\\
\hline
$\kappa_\star$
&
 $-0.57 \begin{array}{ll} {}_{+1.4} & {}_{+2.5} \\ {}^{-1.1} & {}^{-2.9} \end{array}$
&
 $-0.51 \begin{array}{ll} {}_{+1.2} & {}_{+2.6} \\ {}^{-1} & {}^{-2.5} \end{array}$
&
 $-0.52 \begin{array}{ll} {}_{+1.3} & {}_{+2.6} \\ {}^{-1.1} & {}^{-2.5} \end{array}$
&
\\
    \end{tabular}
  \end{footnotesize}
  \caption{
    Constraints on $\Lambda$CDM with isocurvature using Planck 2015 data (P) 
    alone.  The first column uses only the TT data in order to test for the 
    effects of $T \rightarrow E$ leakage on parameter constraints..
    For each parameter, the mean value as well as $68\%$ and $95\%$ upper 
    and lower bounds are shown.
    In some cases, both lower bounds on $Q_n$ are equal due to the prior
    $Q_n \geq 0$, implying that our results only provide an upper bound. 
    \label{t:constraints1d_cmb}
  }
\end{table*}


%% file: table3_cmb+gal.tex

\begin{table*}[tb]
  \tabcolsep=0.02cm
  \begin{scriptsize}
    \begin{tabular}{r||c|c|c|c|c}
      &
      KK+$b_5$, PB
      &
      KK+$\omega_\nu$+$b_5$, PB
      &
      KK, PB
      &
      NB, PB
      &
      PWR, PB
      \\
      
      \hline
\hline
$n_s$
&
 $0.9653 \begin{array}{ll} {}_{+0.0041} & {}_{+0.0081} \\ {}^{-0.0042} & {}^{-0.0086} \end{array}$
&
 $0.9668 \begin{array}{ll} {}_{+0.0048} & {}_{+0.0095} \\ {}^{-0.0047} & {}^{-0.01} \end{array}$
&
 $0.965 \begin{array}{ll} {}_{+0.0042} & {}_{+0.0085} \\ {}^{-0.0045} & {}^{-0.0084} \end{array}$
&
 $0.9651 \begin{array}{ll} {}_{+0.0041} & {}_{+0.0086} \\ {}^{-0.0043} & {}^{-0.0086} \end{array}$
&
 $0.9639 \begin{array}{ll} {}_{+0.0043} & {}_{+0.0089} \\ {}^{-0.0044} & {}^{-0.0086} \end{array}$
\\
\hline
$\sigma_8$
&
 $0.821 \begin{array}{ll} {}_{+0.018} & {}_{+0.034} \\ {}^{-0.017} & {}^{-0.035} \end{array}$
&
 $0.809 \begin{array}{ll} {}_{+0.021} & {}_{+0.036} \\ {}^{-0.015} & {}^{-0.036} \end{array}$
&
 $0.818 \begin{array}{ll} {}_{+0.018} & {}_{+0.034} \\ {}^{-0.018} & {}^{-0.034} \end{array}$
&
 $0.819 \begin{array}{ll} {}_{+0.018} & {}_{+0.033} \\ {}^{-0.018} & {}^{-0.037} \end{array}$
&
 $0.814 \begin{array}{ll} {}_{+0.016} & {}_{+0.032} \\ {}^{-0.02} & {}^{-0.032} \end{array}$
\\
\hline
$h$
&
 $0.6768 \begin{array}{ll} {}_{+0.0052} & {}_{+0.0093} \\ {}^{-0.0045} & {}^{-0.0098} \end{array}$
&
 $0.6717 \begin{array}{ll} {}_{+0.0069} & {}_{+0.012} \\ {}^{-0.0057} & {}^{-0.013} \end{array}$
&
 $0.6764 \begin{array}{ll} {}_{+0.0048} & {}_{+0.0099} \\ {}^{-0.005} & {}^{-0.0098} \end{array}$
&
 $0.6762 \begin{array}{ll} {}_{+0.005} & {}_{+0.01} \\ {}^{-0.005} & {}^{-0.01} \end{array}$
&
 $0.6767 \begin{array}{ll} {}_{+0.0047} & {}_{+0.01} \\ {}^{-0.0052} & {}^{-0.0095} \end{array}$
\\
\hline
$\omega_\mathrm{c}$
&
 $0.1188 \begin{array}{ll} {}_{+0.001} & {}_{+0.0021} \\ {}^{-0.0011} & {}^{-0.002} \end{array}$
&
 $0.1183 \begin{array}{ll} {}_{+0.0011} & {}_{+0.0029} \\ {}^{-0.0014} & {}^{-0.0025} \end{array}$
&
 $0.1189 \begin{array}{ll} {}_{+0.0011} & {}_{+0.0021} \\ {}^{-0.0011} & {}^{-0.0022} \end{array}$
&
 $0.119 \begin{array}{ll} {}_{+0.0011} & {}_{+0.0022} \\ {}^{-0.0012} & {}^{-0.0022} \end{array}$
&
 $0.1189 \begin{array}{ll} {}_{+0.0012} & {}_{+0.0021} \\ {}^{-0.0011} & {}^{-0.0023} \end{array}$
\\
\hline
$\omega_\mathrm{b}$
&
 $0.02226 \begin{array}{ll} {}_{+0.00013} & {}_{+0.00028} \\ {}^{-0.00014} & {}^{-0.00028} \end{array}$
&
 $0.0223 \begin{array}{ll} {}_{+0.00015} & {}_{+0.0003} \\ {}^{-0.00016} & {}^{-0.0003} \end{array}$
&
 $0.02227 \begin{array}{ll} {}_{+0.00014} & {}_{+0.00029} \\ {}^{-0.00014} & {}^{-0.00028} \end{array}$
&
 $0.02225 \begin{array}{ll} {}_{+0.00013} & {}_{+0.00028} \\ {}^{-0.00015} & {}^{-0.00028} \end{array}$
&
 $0.02226 \begin{array}{ll} {}_{+0.00014} & {}_{+0.00029} \\ {}^{-0.00015} & {}^{-0.00028} \end{array}$
\\
\hline
$\omega_\nu$
&
&
 $0.0014 \begin{array}{ll} {}_{+0.0005} & {}_{+0.0016} \\ {}^{-0.0011} & {}^{-0.0014} \end{array}$
&
&
&
\\
\hline
$\tau$
&
 $0.071 \begin{array}{ll} {}_{+0.024} & {}_{+0.043} \\ {}^{-0.02} & {}^{-0.045} \end{array}$
&
 $0.082 \begin{array}{ll} {}_{+0.032} & {}_{+0.049} \\ {}^{-0.027} & {}^{-0.058} \end{array}$
&
 $0.067 \begin{array}{ll} {}_{+0.025} & {}_{+0.043} \\ {}^{-0.022} & {}^{-0.051} \end{array}$
&
 $0.067 \begin{array}{ll} {}_{+0.025} & {}_{+0.044} \\ {}^{-0.024} & {}^{-0.047} \end{array}$
&
 $0.054 \begin{array}{ll} {}_{+0.021} & {}_{+0.042} \\ {}^{-0.03} & {}^{-0.044} \end{array}$
\\
\hline
$Q_n$
&
 $1.0 \begin{array}{ll} {}_{+0.3} & {}_{+1.3} \\ {}^{-1.0} & {}^{-1.0} \end{array}$
&
 $1.1 \begin{array}{ll} {}_{+0.3} & {}_{+1.5} \\ {}^{-1.0} & {}^{-1.1} \end{array}$
&
 $0.96 \begin{array}{ll} {}_{+0.32} & {}_{+1.3} \\ {}^{-0.93} & {}^{-0.96} \end{array}$
&
 $1.1 \begin{array}{ll} {}_{+0.3} & {}_{+1.4} \\ {}^{-1.0} & {}^{-1.1} \end{array}$
&
 $0.012 \begin{array}{ll} {}_{+0.005} & {}_{+0.012} \\ {}^{-0.009} & {}^{-0.012} \end{array}$
\\
\hline
$n_I$
&
 $2.72 \begin{array}{ll} {}_{+1.2} & {}_{+1.2} \\ {}^{-0.69} & {}^{-1.2} \end{array}$
&
 $2.78 \begin{array}{ll} {}_{+1.1} & {}_{+1.2} \\ {}^{-0.59} & {}^{-1.2} \end{array}$
&
 $2.76 \begin{array}{ll} {}_{+1.1} & {}_{+1.2} \\ {}^{-0.59} & {}^{-1.2} \end{array}$
&
 $2.65 \begin{array}{ll} {}_{+0.75} & {}_{+1.2} \\ {}^{-0.7} & {}^{-1.2} \end{array}$
&
 $2.65 \begin{array}{ll} {}_{+0.69} & {}_{+0.92} \\ {}^{-0.4} & {}^{-1.1} \end{array}$
\\
\hline
$\kappa_\star$
&
 $-0.37 \begin{array}{ll} {}_{+1.5} & {}_{+2.6} \\ {}^{-0.98} & {}^{-2.7} \end{array}$
&
 $-0.21 \begin{array}{ll} {}_{+1.3} & {}_{+2.5} \\ {}^{-1.1} & {}^{-2.5} \end{array}$
&
 $-0.21 \begin{array}{ll} {}_{+1.5} & {}_{+2.5} \\ {}^{-1.1} & {}^{-2.4} \end{array}$
&
 $-0.31 \begin{array}{ll} {}_{+1.5} & {}_{+2.6} \\ {}^{-1.2} & {}^{-2.4} \end{array}$
&
\\
    \end{tabular}
  \end{scriptsize}
  \caption{
    Constraints on $\Lambda$CDM with isocurvature using Planck 2015 (P) 
    and BOSS DR11 (B) data.  The first column analyzes the KK model using 
    two extra scale-dependent bias parameters in order to test the
    robustness of our constraints, and the second column varies the sum
    of neutrino masses $\sum m_\nu = 93.14 \omega_\nu$~eV as well as
    these extra bias parameters.
    For each parameter, the mean value as well as $68\%$ and $95\%$ upper 
    and lower bounds are shown.
    In some cases, both lower bounds on $Q_n$ are equal due to the prior
    $Q_n \geq 0$, implying that our results only provide an upper bound.      
    \label{t:constraints1d_cmb+gal}
  }
\end{table*}


%% file: table4_lamppost.tex

\begin{table*}[tb]
  \tabcolsep=0.02cm
  \begin{scriptsize}
    \begin{tabular}{r||c|c|c|c|c}
      &
      LAMP-1, PB
      &
      LAMP-2, PB
      &
      LAMP-10, PB
      &
      BLUE, PB
      &
      HI-BLUE, PB
      \\
      
      \hline
\hline
$n_s$
&
 $0.9653 \begin{array}{ll} {}_{+0.0044} & {}_{+0.0086} \\ {}^{-0.0044} & {}^{-0.0085} \end{array}$
&
 $0.9651 \begin{array}{ll} {}_{+0.0041} & {}_{+0.0086} \\ {}^{-0.0044} & {}^{-0.0083} \end{array}$
&
 $0.9637 \begin{array}{ll} {}_{+0.0039} & {}_{+0.0084} \\ {}^{-0.0045} & {}^{-0.0084} \end{array}$
&
 $0.9649 \begin{array}{ll} {}_{+0.0041} & {}_{+0.0085} \\ {}^{-0.0044} & {}^{-0.0085} \end{array}$
&
 $0.962 \begin{array}{ll} {}_{+0.0045} & {}_{+0.0078} \\ {}^{-0.0038} & {}^{-0.0082} \end{array}$
\\
\hline
$\sigma_8$
&
 $0.819 \begin{array}{ll} {}_{+0.019} & {}_{+0.033} \\ {}^{-0.018} & {}^{-0.035} \end{array}$
&
 $0.817 \begin{array}{ll} {}_{+0.017} & {}_{+0.034} \\ {}^{-0.019} & {}^{-0.034} \end{array}$
&
 $0.814 \begin{array}{ll} {}_{+0.015} & {}_{+0.033} \\ {}^{-0.019} & {}^{-0.033} \end{array}$
&
 $0.819 \begin{array}{ll} {}_{+0.019} & {}_{+0.034} \\ {}^{-0.019} & {}^{-0.038} \end{array}$
&
 $0.812 \begin{array}{ll} {}_{+0.014} & {}_{+0.032} \\ {}^{-0.018} & {}^{-0.027} \end{array}$
\\
\hline
$h$
&
 $0.6761 \begin{array}{ll} {}_{+0.0047} & {}_{+0.0097} \\ {}^{-0.0051} & {}^{-0.0096} \end{array}$
&
 $0.6766 \begin{array}{ll} {}_{+0.0051} & {}_{+0.01} \\ {}^{-0.0051} & {}^{-0.0099} \end{array}$
&
 $0.6765 \begin{array}{ll} {}_{+0.0047} & {}_{+0.01} \\ {}^{-0.0051} & {}^{-0.0096} \end{array}$
&
 $0.6762 \begin{array}{ll} {}_{+0.0049} & {}_{+0.0099} \\ {}^{-0.0048} & {}^{-0.0096} \end{array}$
&
 $0.6771 \begin{array}{ll} {}_{+0.0047} & {}_{+0.0098} \\ {}^{-0.0052} & {}^{-0.0095} \end{array}$
\\
\hline
$\omega_\mathrm{c}$
&
 $0.119 \begin{array}{ll} {}_{+0.0011} & {}_{+0.0022} \\ {}^{-0.0011} & {}^{-0.0022} \end{array}$
&
 $0.1189 \begin{array}{ll} {}_{+0.0011} & {}_{+0.0022} \\ {}^{-0.0011} & {}^{-0.0022} \end{array}$
&
 $0.1189 \begin{array}{ll} {}_{+0.0012} & {}_{+0.0021} \\ {}^{-0.001} & {}^{-0.0022} \end{array}$
&
 $0.119 \begin{array}{ll} {}_{+0.001} & {}_{+0.0021} \\ {}^{-0.0011} & {}^{-0.0021} \end{array}$
&
 $0.1188 \begin{array}{ll} {}_{+0.001} & {}_{+0.0022} \\ {}^{-0.0012} & {}^{-0.0021} \end{array}$
\\
\hline
$\omega_\mathrm{b}$
&
 $0.02226 \begin{array}{ll} {}_{+0.00014} & {}_{+0.00028} \\ {}^{-0.00014} & {}^{-0.00028} \end{array}$
&
 $0.02228 \begin{array}{ll} {}_{+0.00014} & {}_{+0.00029} \\ {}^{-0.00014} & {}^{-0.00028} \end{array}$
&
 $0.02228 \begin{array}{ll} {}_{+0.00014} & {}_{+0.0003} \\ {}^{-0.00015} & {}^{-0.00029} \end{array}$
&
 $0.02226 \begin{array}{ll} {}_{+0.00015} & {}_{+0.0003} \\ {}^{-0.00015} & {}^{-0.00029} \end{array}$
&
 $0.02227 \begin{array}{ll} {}_{+0.00016} & {}_{+0.00028} \\ {}^{-0.00013} & {}^{-0.00028} \end{array}$
\\
\hline
$\tau$
&
 $0.068 \begin{array}{ll} {}_{+0.024} & {}_{+0.044} \\ {}^{-0.023} & {}^{-0.044} \end{array}$
&
 $0.065 \begin{array}{ll} {}_{+0.023} & {}_{+0.043} \\ {}^{-0.024} & {}^{-0.046} \end{array}$
&
 $0.056 \begin{array}{ll} {}_{+0.021} & {}_{+0.041} \\ {}^{-0.028} & {}^{-0.046} \end{array}$
&
 $0.067 \begin{array}{ll} {}_{+0.025} & {}_{+0.044} \\ {}^{-0.024} & {}^{-0.049} \end{array}$
&
 $0.046 \begin{array}{ll} {}_{+0.018} & {}_{+0.04} \\ {}^{-0.032} & {}^{-0.036} \end{array}$
\\
\hline
$Q_n$
&
 $1.4 \begin{array}{ll} {}_{+0.7} & {}_{+1.4} \\ {}^{-1.0} & {}^{-1.4} \end{array}$
&
 $0.93 \begin{array}{ll} {}_{+0.42} & {}_{+0.85} \\ {}^{-0.58} & {}^{-0.93} \end{array}$
&
 $0.19 \begin{array}{ll} {}_{+0.062} & {}_{+0.2} \\ {}^{-0.15} & {}^{-0.19} \end{array}$
&
 $1.1 \begin{array}{ll} {}_{+0.3} & {}_{+1.3} \\ {}^{-1.0} & {}^{-1.1} \end{array}$
&
 $0.062 \begin{array}{ll} {}_{+0.026} & {}_{+0.049} \\ {}^{-0.03} & {}^{-0.052} \end{array}$
\\
\hline
$n_I$
&
 $2.75 \begin{array}{ll} {}_{+0.8} & {}_{+1.2} \\ {}^{-0.67} & {}^{-1.1} \end{array}$
&
 $2.77 \begin{array}{ll} {}_{+0.85} & {}_{+1.2} \\ {}^{-0.61} & {}^{-1.1} \end{array}$
&
 $2.75 \begin{array}{ll} {}_{+0.77} & {}_{+1.2} \\ {}^{-0.65} & {}^{-1.1} \end{array}$
&
&
\\
\hline
$\kappa_\star$
&
&
&
&
 $-0.45 \begin{array}{ll} {}_{+1.2} & {}_{+2.2} \\ {}^{-0.89} & {}^{-2.2} \end{array}$
&
\\
    \end{tabular}
  \end{scriptsize}
  \caption{
    Constraints on lamppost models using Planck and BOSS data.  LAMP-$N$ 
    is KK with $\kappa_\star = \ln(N)$, BLUE is KK with $n_I = 3.9$, and 
    HI-BLUE is BLUE with $\kappa_\star = \ln(10)$.
    For each parameter, the mean value as well as $68\%$ and $95\%$ upper 
    and lower bounds are shown.
    In some cases, both lower bounds on $Q_n$ are equal due to the prior
    $Q_n \geq 0$, implying that our results only provide an upper bound.      
    \label{t:constraints1d_lamppost}
  }
\end{table*}


%% file: table5_blue_hiblue.tex

\begin{table*}[tb]
  \tabcolsep=0.02cm
  \begin{footnotesize}
    \begin{tabular}{r||c|c|c|c}
      &
      BLUE, P
      &
      HI-BLUE, P
      &
      BLUE, PB
      &
      HI-BLUE, PB
      \\
      
      \hline
\hline
$n_s$
&
 $0.9657 \begin{array}{ll} {}_{+0.0051} & {}_{+0.01} \\ {}^{-0.0052} & {}^{-0.01} \end{array}$
&
 $0.9628 \begin{array}{ll} {}_{+0.0046} & {}_{+0.0099} \\ {}^{-0.0048} & {}^{-0.01} \end{array}$
&
 $0.9649 \begin{array}{ll} {}_{+0.0041} & {}_{+0.0085} \\ {}^{-0.0044} & {}^{-0.0085} \end{array}$
&
 $0.962 \begin{array}{ll} {}_{+0.0045} & {}_{+0.0078} \\ {}^{-0.0038} & {}^{-0.0082} \end{array}$
\\
\hline
$\sigma_8$
&
 $0.844 \begin{array}{ll} {}_{+0.019} & {}_{+0.035} \\ {}^{-0.017} & {}^{-0.037} \end{array}$
&
 $0.838 \begin{array}{ll} {}_{+0.022} & {}_{+0.035} \\ {}^{-0.02} & {}^{-0.042} \end{array}$
&
 $0.819 \begin{array}{ll} {}_{+0.019} & {}_{+0.034} \\ {}^{-0.019} & {}^{-0.038} \end{array}$
&
 $0.812 \begin{array}{ll} {}_{+0.014} & {}_{+0.032} \\ {}^{-0.018} & {}^{-0.027} \end{array}$
\\
\hline
$h$
&
 $0.6764 \begin{array}{ll} {}_{+0.0069} & {}_{+0.014} \\ {}^{-0.007} & {}^{-0.014} \end{array}$
&
 $0.6767 \begin{array}{ll} {}_{+0.0069} & {}_{+0.014} \\ {}^{-0.0064} & {}^{-0.014} \end{array}$
&
 $0.6762 \begin{array}{ll} {}_{+0.0049} & {}_{+0.0099} \\ {}^{-0.0048} & {}^{-0.0096} \end{array}$
&
 $0.6771 \begin{array}{ll} {}_{+0.0047} & {}_{+0.0098} \\ {}^{-0.0052} & {}^{-0.0095} \end{array}$
\\
\hline
$\omega_\mathrm{c}$
&
 $0.119 \begin{array}{ll} {}_{+0.0015} & {}_{+0.0031} \\ {}^{-0.0016} & {}^{-0.0031} \end{array}$
&
 $0.1189 \begin{array}{ll} {}_{+0.0014} & {}_{+0.003} \\ {}^{-0.0015} & {}^{-0.003} \end{array}$
&
 $0.119 \begin{array}{ll} {}_{+0.001} & {}_{+0.0021} \\ {}^{-0.0011} & {}^{-0.0021} \end{array}$
&
 $0.1188 \begin{array}{ll} {}_{+0.001} & {}_{+0.0022} \\ {}^{-0.0012} & {}^{-0.0021} \end{array}$
\\
\hline
$\omega_\mathrm{b}$
&
 $0.02228 \begin{array}{ll} {}_{+0.00017} & {}_{+0.00033} \\ {}^{-0.00016} & {}^{-0.00033} \end{array}$
&
 $0.0223 \begin{array}{ll} {}_{+0.00017} & {}_{+0.00032} \\ {}^{-0.00016} & {}^{-0.00032} \end{array}$
&
 $0.02226 \begin{array}{ll} {}_{+0.00015} & {}_{+0.0003} \\ {}^{-0.00015} & {}^{-0.00029} \end{array}$
&
 $0.02227 \begin{array}{ll} {}_{+0.00016} & {}_{+0.00028} \\ {}^{-0.00013} & {}^{-0.00028} \end{array}$
\\
\hline
$\tau$
&
 $0.098 \begin{array}{ll} {}_{+0.026} & {}_{+0.048} \\ {}^{-0.023} & {}^{-0.051} \end{array}$
&
 $0.081 \begin{array}{ll} {}_{+0.03} & {}_{+0.05} \\ {}^{-0.026} & {}^{-0.057} \end{array}$
&
 $0.067 \begin{array}{ll} {}_{+0.025} & {}_{+0.044} \\ {}^{-0.024} & {}^{-0.049} \end{array}$
&
 $0.046 \begin{array}{ll} {}_{+0.018} & {}_{+0.04} \\ {}^{-0.032} & {}^{-0.036} \end{array}$
\\
\hline
$Q_n$
&
 $1.1 \begin{array}{ll} {}_{+0.3} & {}_{+1.5} \\ {}^{-1.1} & {}^{-1.1} \end{array}$
&
 $0.047 \begin{array}{ll} {}_{+0.024} & {}_{+0.044} \\ {}^{-0.028} & {}^{-0.047} \end{array}$
&
 $1.1 \begin{array}{ll} {}_{+0.31} & {}_{+1.3} \\ {}^{-1} & {}^{-1.1} \end{array}$
&
 $0.062 \begin{array}{ll} {}_{+0.026} & {}_{+0.049} \\ {}^{-0.03} & {}^{-0.052} \end{array}$
\\
\hline
$\kappa_\star$
&
 $-0.68 \begin{array}{ll} {}_{+0.99} & {}_{+2.2} \\ {}^{-0.79} & {}^{-2.2} \end{array}$
&
&
 $-0.45 \begin{array}{ll} {}_{+1.2} & {}_{+2.2} \\ {}^{-0.89} & {}^{-2.2} \end{array}$
&
\\
    \end{tabular}
  \end{footnotesize}
  \caption{
    Constraints on extremely blue-tilted lamppost models with $n_I = 3.9$.
    BLUE allows $\kappa_\star$ to vary while HI-BLUE fixes it at $\ln(10)$.
    Constraints are from Planck data alone (P) or Planck plus BOSS (PB).
    For each parameter, the mean value as well as $68\%$ and $95\%$ upper 
    and lower bounds are shown.
    In some cases, both lower bounds on $Q_n$ are equal due to the prior
    $Q_n \geq 0$, implying that our results only provide an upper bound.      
    \label{t:constraints1d_blue_hiblue}
  }
\end{table*}
